\documentclass[aip,amsmath,amssymb,reprint]{revtex4-1}

\usepackage{graphicx}
\usepackage{dcolumn}
\usepackage{bm}

\usepackage[utf8]{inputenc}
\usepackage[T1]{fontenc}
\usepackage{mathptmx}
\usepackage{etoolbox}

\makeatletter
\def\@email#1#2{%
 \endgroup
 \patchcmd{\titleblock@produce}
  {\frontmatter@RRAPformat}
  {\frontmatter@RRAPformat{\produce@RRAP{*#1\href{mailto:#2}{#2}}}\frontmatter@RRAPformat}
  {}{}
}%
\makeatother
\begin{document}


\title[Dissociation of indene]{Dissociation and radiative stabilization of the indene cation:\\ The nature of the C-H bond and astrochemical implications}

\author{M. H. Stockett}
\affiliation{Department of Physics, Stockholm University, SE-10691 Stockholm, Sweden}
 \email{Mark.Stockett@fysik.su.se}
\author{A. Subramani}
\affiliation{Department of Physics, Stockholm University, SE-10691 Stockholm, Sweden}
\author{C. Liu}
\affiliation{School of Chemistry, University of Melbourne, Parkville, VIC 3010, Australia}
\author{S. J. P. Marlton}
\affiliation{School of Chemistry, University of Melbourne, Parkville, VIC 3010, Australia}
\author{E. K. Ashworth}
\affiliation{Chemistry, Faculty of Science, University of East Anglia, Norwich NR4 7TJ, United Kingdom}
\author{H. Cederquist}
\affiliation{Department of Physics, Stockholm University, SE-10691 Stockholm, Sweden}
\author{H. Zettergren}
\affiliation{Department of Physics, Stockholm University, SE-10691 Stockholm, Sweden}
\author{J. N. Bull}%
\affiliation{Chemistry, Faculty of Science, University of East Anglia, Norwich NR4 7TJ, United Kingdom}
\affiliation{Centre for Photonics and Quantum Science, University of East Anglia, Norwich NR4 7TJ, United Kingdom}

\date{\today}

\begin{abstract}
Indene (C$_9$H$_8$) is the only polycyclic pure hydrocarbon identified in the interstellar medium to date, with an observed abundance orders of magnitude higher than predicted by astrochemical models. The dissociation and radiative stabilization of vibrationally-hot indene cations is investigated by measuring the time-dependent neutral particle emission rate from ions in a cryogenic ion-beam storage ring for up to 100~ms. Time-resolved measurements of the kinetic energy released upon hydrogen atom loss from C$_9$H$_8^+$, analyzed in view of a model of tunneling through a potential energy barrier, provides the dissociation rate coefficient. Master equation simulations of the dissociation in competition with vibrational and electronic radiative cooling reproduce the measured dissociation rate. We find that radiative stabilization arrests one of the main C$_9$H$_8$ destruction channels included in astrochemical models, helping to rationalize its high observed abundance.
\end{abstract}

\maketitle

\section{Introduction}
Polycyclic aromatic hydrocarbons (PAHs) have long been thought to be ubiquitous in the interstellar medium (ISM). This is based on the infrared emission bands observed by astronomers at wavelengths coincident with their characteristic vibrational transition energies \cite{Tielens2008}. However, these bands are common to PAHs as a class and have proven difficult to assign to specific PAH molecules. It was only in the last few years that indene \cite{Burkhardt2021,Cernicharo2021}, 2-cyanoindene \cite{Sita2022}, and two isomers of cyano-naphthalene \cite{McGuire2021} were identified in space in the Taurus molecular cloud, TMC-1 by comparing astronomical microwave spectra with known rotational emission lines. Very recently, several larger cyano-functionalized PAHs have been identified by the same methods \cite{Cernicharo2024,Wenzel2024}. Interestingly, the observed interstellar abundance of indene (C$_9$H$_8$, FIG.~\ref{fig_desiree}), the only polycyclic pure hydrocarbon identified to date, was more than four orders of magnitude higher than predicted by astrochemical modeling \cite{Sita2022}. These models neglect radiative cooling and assume that PAHs are rapidly broken down into linear fragments following ionizing collisions with small cations including C$^+$ and H$_3^+$ \cite{McGuire2021}.

A previous study \cite{Stockett2023} on 1-cyanonaphthalene by some of the present authors found that the energized cation is efficiently stabilized by recurrent fluorescence (RF) -- the emission of optical photons from thermally excited electronic states \cite{Leger1988,Boissel1997}. This rapid radiative cooling closes some of the destruction pathways included in astrochemical models that underpredicted the abundanceof that molecule in TMC-1 by six orders of magnitude \cite{McGuire2021}. Numerous laboratory studies have confirmed that RF, rather than infrared photon emission from vibrational relaxation, is the primary radiative stabilization mechanism for PAH cations \cite{Martin2013,Martin2015,Ji2017,Saito2020,Stockett2020b,Stockett2023,Lee2023,NavarroNavarrete2023,Bernard2023}, with few exceptions \cite{Zhu2022}.

We present a study of the unimolecular dissociation and radiative stabilization of the indene radical cation, C$_9$H$_8^+$, Ind$^+$. The dissociation rate coefficient and activation energy are determined from time-resolved measurements of kinetic energy release distributions of an ensemble of vibrationally excited ions isolated in a cryogenic ion-beam storage ring. RF rate coefficients are calculated based on \textit{ab initio} molecular dynamics simulations.  Master equation modeling including dissociation, RF, and infrared vibrational radiative cooling processes closely reproduce the measured absolute dissociation rate of the stored ion ensemble. We find that Ind$^+$, in contrast to fully $sp^2$-hybridized PAH cations, cools mainly by IR photon emission, with at most a small contribution from RF. Nevertheless, we find that Ind$^+$ can be expected to be radiatively stabilized following charge-exchange reactions between C$^+$ and neutral indene, one of the most important destruction channels for PAHs included in astrochemical models of molecular clouds.

\section{Methods}

\subsection{Experiments}
\label{sec_exp}
\begin{figure*}
\includegraphics[width=\textwidth]{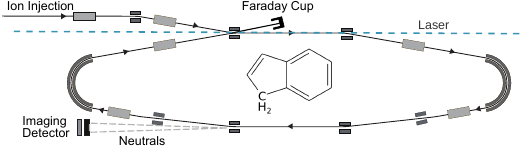}
\caption{DESIREE electrostatic ion-beam storage ring. Neutral fragments formed in the lower straight section are not constrained by the storage ring's electric fields and impact on a detector. Inset shows the structure of indene noting the $sp^3$ functional group.}
\label{fig_desiree}
\end{figure*}

Experiments were conducted using the DESIREE (Double ElectroStatic Ion Ring ExpEriment) cryogenic ion-beam storage ring infrastructure located at Stockholm University \cite{Thomas2011,Schmidt2013} using procedures similar to earlier experiments on cyano-naphthalene \cite{Stockett2023}. Indene (Sigma-Aldrich, $>$99\%) vapor was introduced into a electron cyclotron resonance (ECR, Panteknik Monogan) ion source. The ECR source produces molecular ions with high internal energies. Positive ions were extracted from the source, accelerated to 86~keV, and those with $m/z=116$ were selected with a bending magnet. Ion beams were stored using electrostatic deflectors in the storage ring depicted in FIG.~\ref{fig_desiree}. Neutral fragments emitted from vibrationally hot ions in the lower straight section of the ring (see FIG~\ref{fig_desiree}) impinged on a 75~mm diamater Z-stack microchannel plate detector (Photonis) with a phosphor screen anode (phosphor type P24). The position and time of each detected fragment was recorded using a TimePix3 hybrid pixel detector (Amsterdam Scientific) located outside of the DESIREE vacuum chamber. At the end of each storage cycle, the beam was dumped into a Faraday cup coupled to a fast amplifier for current normalization.

To analyze the kinetic energy released during dissociation, the event data from the TimePix was first binned according to time after ionization, with a bin width equal to the revolution period of the ions around the ring. Two dimensional histograms of the detector images for each revolution were centered and azimuthally integrated using the PyAbel package \cite{pyabel}. The resulting time-dependent radial intensity distributions were then further averaged over time windows with widths linearly increasing with storage time, to improve signal-to-noise at late times where the count rate is low. The radial intensity distribution includes contributions from neutral fragments emitted along the straight section of the storage ring and projected onto the detector plane (see FIG.~\ref{fig_desiree}). We fit analytic kinetic energy release (KER) distribution models to this final projected distribution by first applying a matrix transformation from energy to radial distance considering, for each energy bin, the range of radii over which the differential intensity will be spread given the finite length of the straight section. The projection of this spherically symmetric distribution onto the detector plane is achieved with a one-dimensional forward Abel transform. For more information on the analysis, see the Appendix.

While our numerical analysis of the KER is based on fits to the detector-plane radial intensity distribution, selected KER distributions are shown in Sec.~\ref{sec_results} to facilitate comparison to the literature. To account for the numerical error introduced by the inverse Abel transform, we invert and average 128 radial distributions with intensities normally distributed according to the measured distribution and its counting statistics. The resulting distribution is calibrated assuming all neutrals are emitted from the mid-point of the straight section, \textit{i.e.}:
\begin{equation}
\epsilon(\rho) = \frac{m_{neut}}{m_{cat}}E_{acc}\left(\frac{\rho}{L_{mid}}\right)^2
\label{eq_epsilon}
\end{equation}
where $\epsilon(\rho)$ is the KER associated with the radius $\rho$, $m_{neut}$ and $m_{cat}$ are the masses of the neutral and cationic reaction products, $E_{acc}=86$~keV is the beam energy, and $L_{mid}=1.7$~m is the distance from the detector to the mid-point of the straight section. This leads to some distortion of the plotted KER distribution (see Appendix).

A photo-activation technique was employed to probe the evolution of the internal energy distribution on longer timescales, after the initially hot ions have all either dissociated or been radiatively stabilized. In these experiments, light from a wavelenght-tunable optical parametric oscillator laser system was overlapped colinearly with the stored ion beam as shown in FIG~\ref{fig_desiree}. Following rapid internal conversion and intramolecular vibrational redistribution, photo-activation increases the vibrational energy of the excited ions by a known, fixed amount (the photon energy). Photo-activated ions are thus re-heated to energies above the dissociation threshold and neutral fragments are again emitted and detected using the same procedures detailed above for the source-heated ions. 

\subsection{Quantum chemical calculations}

The vibrational frequencies and infrared intensities of Ind$^+$ were calculated at the LC-$\omega$HPBE/cc-pVTZ level in Gaussian \cite{g16}. The infrared emission rate coefficients were computed using the simple harmonic cascade (SHC) approximation \cite{Chandrasekaran2014}:

\begin{equation}
k^{IR}_s=A_s^{IR}\sum_{v=1}^{v\leq E/h\nu_s} \frac{\rho(E-vh\nu_s)}{\rho(E)},
\label{eq_kir}
\end{equation}
where $v$ is the vibrational quantum number, and $h\nu_s$ and $A_s$ are the transition energy and Einstein coefficient of vibrational mode $s$, respectively. The density of vibrational states $\rho(E)$ is calculated using the Beyer-Swineheart algorithm \cite{Beyer1973}. The SHC model considers only $v\rightarrow v-1$ fundamental transitions, and the transition energies and level densities are assumed to be harmonic. However, we use anharmonic (VPT2) fundamental frequencies and intensities in our simulations, as these have been shown to better reproduce experiments \cite{Bull2019a}.

The RF emission rate coefficients are parameterized as \cite{Boissel1997}:

\begin{equation}
k^{RF}_j(E)=A^{RF}_j\frac{\rho(E-E_j)}{\rho(E)},
\label{eq_krf}
\end{equation}
where $E_j$ and $A^{RF}_j$ are the transition energy and Einstein coefficient of the D$_j\leftarrow$D$_0$ electronic transition. 

Quantitative studies, by the present collaboration, as well as others, have generally found that Eq.~\ref{eq_krf} drastically underestimates RF cooling rates when the parameters $E_j$ and $A^{RF}_j$ are estimated using quantum chemical methods such as TD-DFT \cite{Martin2013,Stockett2020b}. It was previously reported that this discrepancy could be narrowed by consideration of Herzberg-Teller (HT) vibronic coupling, which greatly increases the transition probability of the lowest-energy optical transition of PAHs \cite{Stockett2023}. HT simulations of open-shell systems remain a challenge for computational chemistry programs, and recently a more generally applicable method based on molecular dynamics simulations was introduced \cite{Bull2025}.

Here, we performed temperature-dependent \textit{ab initio} molecular dynamics (AIMD) \cite{Iftimie2005} at the $\omega$B97X-D/cc-pVTZ level of theory \cite{Dunning1989,Chai2008}, with canonical sampling through velocity rescaling \cite{Bussi2007}. Trajectories at given initial temperatures, converted to vibrational energies according to the caloric curve based on the state density, were run for 1~ps in 0.5~fs steps, with the vertical transition energies and oscillator strengths for the D$_j\leftarrow$~D$_0$ states computed at each step. The RF rate coefficients for fixed thermal energies $E$ were then calculated and time-averaged for each trajectory. 

Dissociation energies for H-loss from the $sp^3$ site were performed using the DSD-PBEP86-D3(BJ)/def2-TZVPP and CCSD(T)/cc-pVTZ theories.

\subsection{Master equation simulations}

Our approach to master equation simulations of PAH cooling dynamics has been described previously \cite{Stockett2020b}. The vibrational energy distribution $g(E,t)$ of the ensemble is initialized as a normalized Boltzmann distribution at $t=0$. The master equation propagates the distribution according to:
\begin{multline}
\frac{d}{dt}g(E,t)=-k^{diss}(E)g(E,t) \\ +\sum_{s} \left[ k^{IR}_{s}(E+h\nu_s)g(E+h\nu_s,t)-k^{IR}_{s}(E)g(E,t)\right] \\ +\sum_{j} \left[ k^{RF}_{j}(E+E_j)g(E+h\nu_j,t)-k^{RF}_{j}(E)g(E,t)\right].
\label{eqn_master}
\end{multline}
The first term gives the depletion of the population by unimolecular dissociation. On the next line, the first term in brackets represents $v+1\rightarrow v$ vibrational emission from levels above $E$, while the second is $v\rightarrow v-1$ emission to levels below $E$. Similarly, the terms on the third line account for RF emission from the electronically excited states D$_j$ ($j=1,2,3$) to the ground state D$_0$. The time step $dt$ is chosen to match the experimental data, with 32 extra points prior to the first experimental time bin to account for ion decays during the transit time from the ion source to the storage ring. The dissociation rate is given by $\Gamma(t)=\int k^{diss}(E)g(E,t)dE$.

\section{Results}
\label{sec_results}

\subsection{Recurrent fluorescence rate coefficients}

\begin{figure}
\includegraphics[width=\columnwidth]{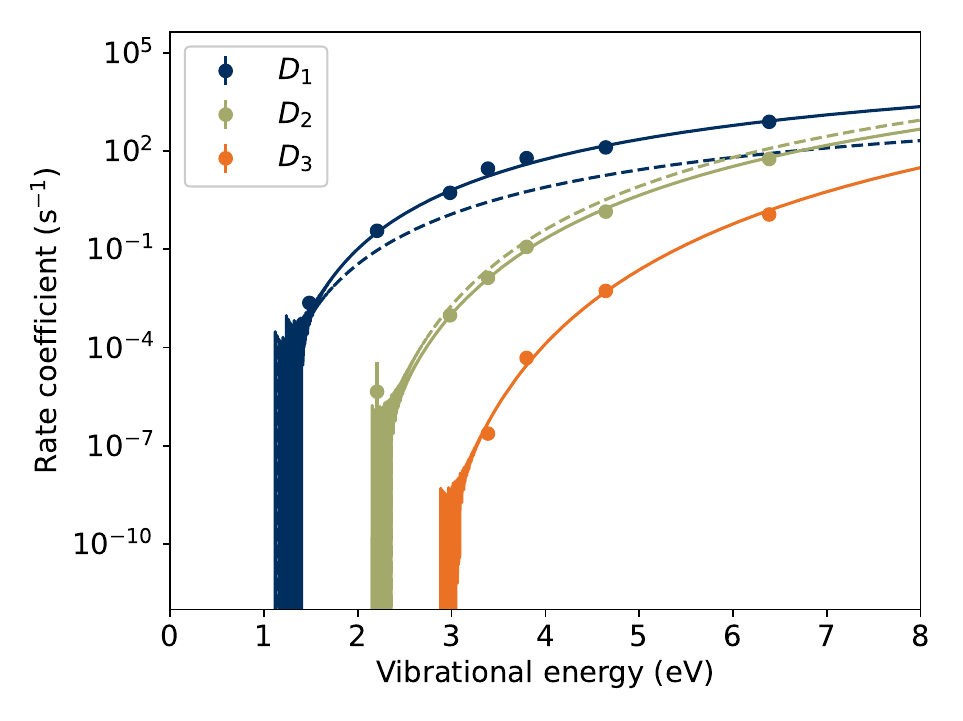}
\caption{RF rate coefficients for the lowest three electronic transitions of Ind$^+$. The points are the time-averaged values from individual AIMD trajectories at fixed internal energies. The solid curves are fits of Eq.~\ref{eq_krf}. The dashed curves for D$_1$ and D$_2$ are calculated rate coefficients using the TD-DFT values from Chalyavi \textit{et al.} \cite{Chalyavi2013}}
\label{fig_rf}
\end{figure}

Our calculated RF rate coefficients, based on our AIMD simulations, are presented in FIG.~\ref{fig_rf}. The solid curves are fits of Eq.~\ref{eq_krf} to the points which are time averaged values from AIMD trajectories with given initial temperatures. The fit parameters $E_j$ and $f_{j0}$ are given in TABLE~\ref{tab_rf}, along with experimental and calculated values \cite{Chalyavi2013} from Chalyavi \textit{et al}. Prior to fitting, the D$_{n}$ state energies in the AIMD simulations were scaled by a factor of 1.03 so that the resulting value of $E_2$ agrees with the 0-0 transition energy Chalyavi \textit{et al.} measured using He-tagging messenger spectroscopy \cite{Chalyavi2013}. Chu \textit{et al.} reported \cite{Chu2023} a nearly identical energy of this transition at 17379.3~cm$^{-1}$ for bare Ind$^+$, while Nagy \textit{et al.} found \cite{Nagy2013} 17249~cm$^{-1}$ in a Ne matrix. Significantly, the oscillator strength of the D$_1\leftarrow$~D$_0$ transition is an order of magnitude greater in our AIMD simulation than in the TD-DFT calculations of Chalyavi \textit{et al.}, leading to a commensurately increased RF rate coefficient. 


\begin{table}
\caption{Transition energies $E_j$ (cm$^{-1}$) and oscillator strengths $f_{j0}$ for D$_j\leftarrow$D$_0$ transitions in Ind$^+$.}
\begin{tabular}{|c|cc|ccc|}
\hline 
 & \multicolumn{2}{c|}{This work (AIMD)} & \multicolumn{3}{c|}{Chalyavi \textit{et al.} \cite{Chalyavi2013}}\\
$j$ & $E_j$ & $f_{j0}$ & $E_j$ exp. & $E_j$ calc. & $f_{j0}$ calc. \\ 
\hline 
1 & 10000(300) & 0.006(1) & & 9026 & 0.0004 \\ 
2 & 17400(100) & 0.030(3) & 17379(15) & 20243 & 0.0563 \\ 
3 & 23300(300) & 0.05(3) & & 26971 & 0.0000 \\ 
\hline 
\end{tabular} 
\label{tab_rf}
\end{table}

\subsection{Kinetic energy release}

\begin{figure}
\includegraphics[width=\columnwidth]{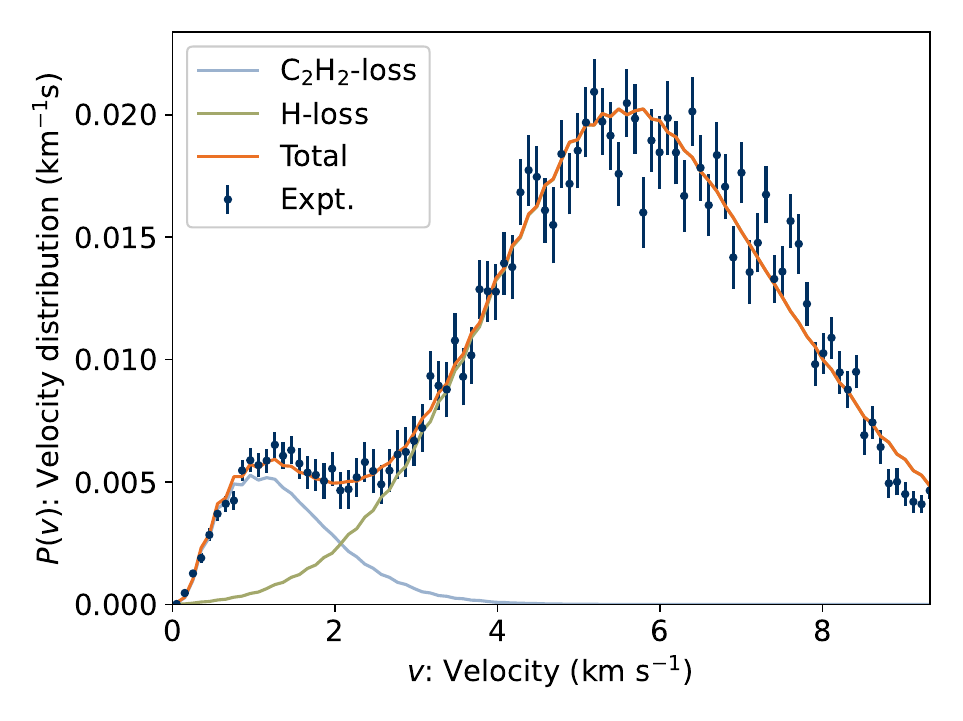}
\caption{Velocity distribution of neutral products emitted from stored ion beam 56-242~$\mu$s after ionization. The smooth curves are fits of Eq.~\ref{eq_kerd}. H-loss is from Ind$^+$ while C$_2$H$_2$-loss is from contaminant [Ind-H]$^+$ with one $^{13}$C atom.}
\label{fig_lumps}
\end{figure}

The velocity distribution of the neutral products, integrated over the first eight revolutions of the stored ions in DESIREE ($t=56-242$~$\mu$s after ionization), is shown in FIG.~\ref{fig_lumps}. The distribution is bimodal with the dominant component with higher mean velocity being due to H-loss from Ind$^+$.  We attribute the low-velocity component to C$_2$H$_2$-loss from $^{13}$C-indenyl ($^{13}$C$^{12}$C$_8$H$_7^+$), which has the same mass-to-charge ratio as our ion of interest ($^{12}$C$_9$H$_8^+$). Owing to its fully $sp^2$-hybridized structure, the indenyl cation [Ind-H]$^+$ has a higher dissociation energy for H-loss than Ind$^+$, and was found to decay exclusively by C$_2$H$_2$-loss in previous experiments at DESIREE \cite{Bull2025}. Given that the dissociation threshold energy for C$_2$H$_2$-loss from Ind$^+$ is significantly higher than for H-loss, we do not expect these channels to be competitive on such long timescales \cite{West2014}. We can also rule out sequential fragmentation processes in which stored Ind$^+$ first lose an H atom to form [Ind-H]$^+$, and then fragments again by C$_2$H$_2$-loss. This is because ions with such high internal energies as to undergo multi-fragmentation would not survive to the point of mass selection by the bending magnet.

\begin{figure}
\includegraphics[width=\columnwidth]{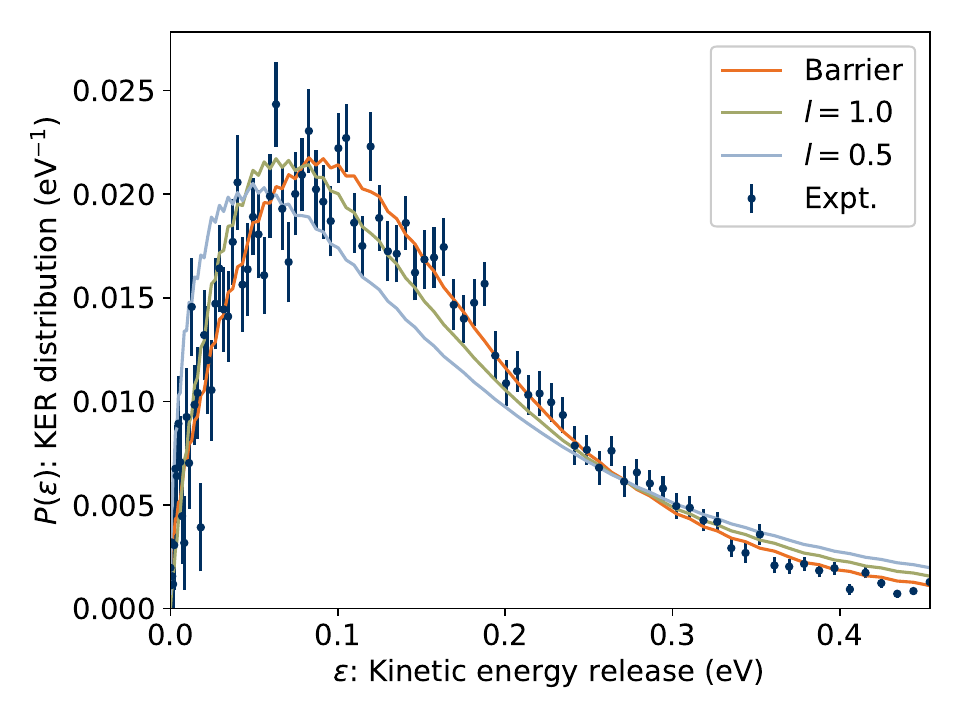}
\caption{KER distribution of Ind$^+$ integrated over the 25--50~ms time window following ionization in an ECR ion source. Solid curves are fits of Eqs.~\ref{eq_kerd} and \ref{eq_mfa}.}
\label{fig_kerd}
\end{figure}

To handle the [Ind-H]$^+$ contamination in our analysis of the KER distributions, a separate measurement was made of all-$^{12}$C [Ind-H]$^+$ under the same experimental conditions, giving the shape parameters needed to fit the combined distributions in FIG.~\ref{fig_lumps}. As the dissociation of [Ind-H]$^+$ is quenched much more rapidly than that of Ind$^+$ (see Sec.~\ref{sec_rate}), the parameters for Ind$^+$ were obtained by integrating the KER distributions over the 25--50~ms time window as shown in FIG.~\ref{fig_kerd}.

The kinetic energy released in H-loss reactions of hydrocarbons has long been known to be anomalously large \cite{Klots1976}, and the shape of the KER distribution is notoriously difficult to model \cite{Gridelet2006}. The reaction generally shows no reverse activation barrier, and indeed no energy gap in the KER distribution is observed. However, the emission of low-energy ($<$0.05~eV) H atoms is notably suppressed, skewing the distribution beyond what can be readily accommodated by otherwise successful models \cite{Klots1976}. In their study of the benzene cation, Gridelet \textit{et al.} argue that the high KER is due to the difference in electronic configurations of the reactant and product states \cite{Gridelet2006}. The nonadiabatic character of the reaction requires passage through a conical intersection to a dissociative state of the appropriate symmetry, which Gridelet \textit{et al.} show can qualitatively be understood as tunneling through a modified centrifugal barrier. 

Our measured KER distributions were fit by the model of Hansen \cite{Hansen2018}, who considered tunneling and reflection from a potential barrier given by an inverted parabola as a model of a transition state. The shape of the KER distribution in this model is given by:
\begin{multline}
P(\epsilon) = \frac{e^{\beta'}}{e^{\beta'}+1}e^{-(\epsilon-\Delta E)/k_BT_p}, \\ \mathrm{where}\ \beta'=4\pi\frac{\Delta E}{\hbar\omega}\left(\sqrt{\frac{\epsilon}{\Delta E}}-1\right)
\label{eq_kerd}
\end{multline}
\noindent where $\epsilon$ is the kinetic energy release, $\Delta E$ is the height of the barrier above the product channel, $T_p$ is the temperature of the decaying ion, and $\hbar\omega$ is related to the curvature of the transition state. The fitted curve labeled `Barrier' in FIG.~\ref{fig_kerd} returned $\Delta E=0.33(8)$~eV, which is large compared to the typical KER. This is in contrast to the case of a reverse reaction barrier where $\Delta E$ is measured as the gap between 0~eV and the onset of the KER distribution \cite{Hansen2018,Laskin2001}.

For comparison, we have included fits of the measured KER distribution to the conventional ``model-free'' expression \cite{Klots1991,Laskin2001}:
\begin{equation}
P(\epsilon)\propto\epsilon^le^{-\epsilon/k_BT_p},
\label{eq_mfa}
\end{equation}
\noindent where the parameter $l$ characterizes the interaction between the products. For the Langevin ion/induced-dipole interaction, $l=0.5$, while $l=1$ corresponds to the limiting case of a hard sphere. Neither agree with the experimental data as well as the Hansen barrier model. The best agreement between the model-free expression and the measured KER distribution (not shown) is obtained when $l$ is allowed to take the non-physical value of $~1.4$, which is associated with the anomalously high KER of H-loss reactions from hydrocarbons \cite{Klots1976,Gridelet2006}.

\begin{figure}
\includegraphics[width=\columnwidth]{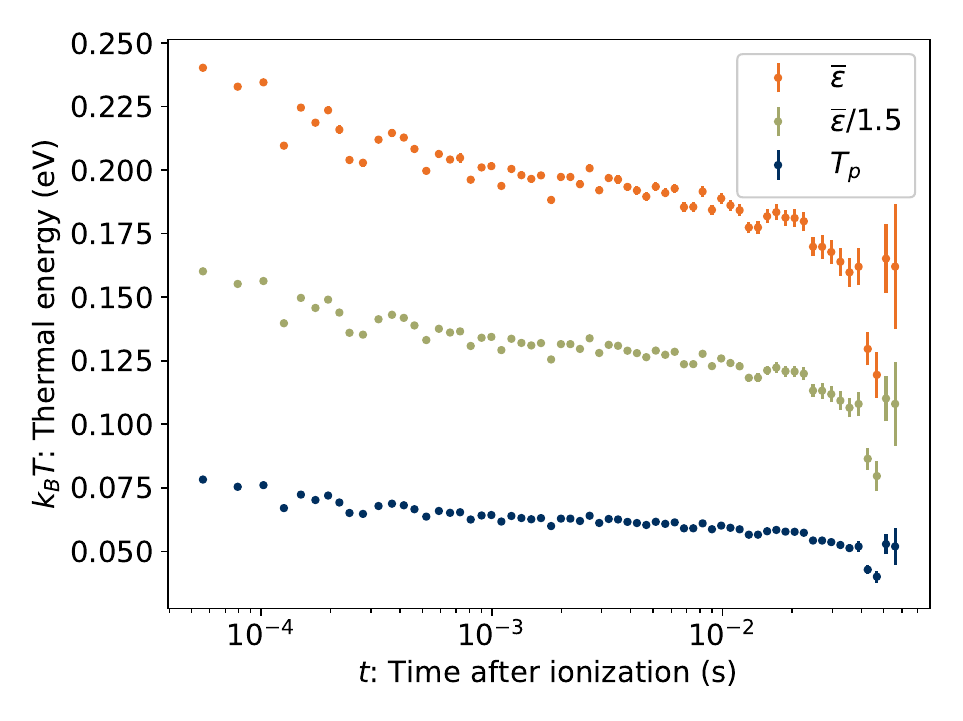}
\caption{Product thermal energies for dissociation of Ind$^+$ obtained from fits of Eq.~\ref{eq_kerd} to measured KER distributions.}
\label{fig_temp}
\end{figure}

The time-dependent temperature of the decaying ions is determined from the analysis of KER distributions. The temperature of an isolated molecule having total vibrational energy $E$ is that of a fictitious Boltzmann distribution with a mean energy \cite{Andersen2001} equal to $E$. One must distinguish between the temperature of the reactant $T_r$, corresponding to a vibrational energy $E$, the product temperature $T_p$, with internal energy $E-E_a$, and the effective emission temperature $T_e$, which reconciles the canonical and microcanonical descriptions of the rate coefficient:
\begin{equation}
k^{diss}=Ae^{-E_a/k_BT_e}=A\frac{\rho(E-E_a)}{\rho(E)}.
\end{equation}
\noindent According to Finite Heat Bath theory \cite{Andersen2002}, $T_e\approx(T_r+T_p)/2$. The KER distribution gives information about $T_p$ \cite{Andersen2002}. In the model-free approach (Eq.~\ref{eq_mfa}), the average KER is a measure of the temperature:
\begin{equation}
\bar{\epsilon}=\int\epsilon P(\epsilon)d\epsilon/\int P(\epsilon)d\epsilon=(l+1)k_BT_p.
\label{eq_ebar}
\end{equation}
\noindent For the Langevin interaction, $l=0.5$, and so $\bar{\epsilon}=1.5k_BT_p$. In the absence of tunneling corrections, these two methods of determining the product temperature usually agree. As shown in FIG.~\ref{fig_temp}, we find $\bar{\epsilon}$ to be roughly a factor of three higher than the temperature returned by a fit of Eq.~\ref{eq_kerd}, portending a non-physical value of $l\approx 2$. 

\subsection{Dissociation rate coefficient}
\label{sec_kdiss}

\begin{figure}
\includegraphics[width=\columnwidth]{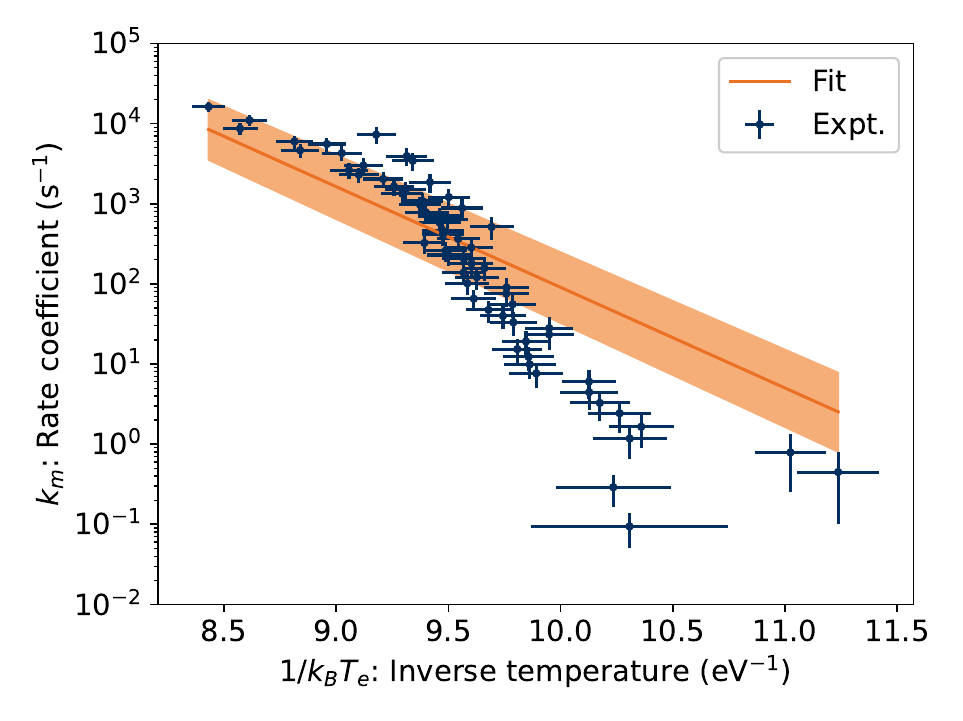}
\caption{Arrhenius plot for H-loss from Ind$^+$. The effective emission temperatures $T_e$ are determined from fits of Eq.~\ref{eq_kerd} to the measured KER distributions, with corrections for the finite heat bath (Eq.~\ref{eq_heatbath}). The most-probable dissociation rate coefficient $k_m$ is determined from the measured count rate using Eq.~\ref{eq_km}. The solid line is a fit of Eq.~\ref{eq_arrh}.}
\label{fig_arrh}
\end{figure}

Proceeding with the values of $T_p$ obtained from Eq.~\ref{eq_kerd}, we determine the Arrhenius parameters of the dissociation rate coefficient from a fit to the measured data of the expression:
\begin{equation}
\log(k_m)=\log(A)-E_a/k_BT_p.
\label{eq_arrh}
\end{equation}
The most probable dissociation rate coefficient $k_m(t)$ observed at time $t$ is given by:
\begin{equation}
k_m(t)=\phi_H(t)R(t)/r_0
\label{eq_km}
\end{equation}
\noindent where $\phi_H(t)$ is the fraction of counts due to H-loss, determined from the fit to the KER distribution, $R(t)$ is the total count rate, and $r_0$ is a dimensionless normalization constant (see Sec.~\ref{sec_rate}). To determine the effective emission temperature $T_e$, the energy of the decaying ions $E_e$, is calculated as part of the fit from the KER distributions using the computed caloric curve $E_{tot}(T_e)$ for the product [Ind-H]$^+$ and the second order finite heat bath correction \cite{Andersen2001}:
\begin{equation}
E_e=E_{tot}(T_p)+\frac{E_a}{2}-\frac{E_a^2}{12(E_{tot}(T_p)+E_a)}.
\label{eq_heatbath}
\end{equation}
From $E_e$, the emission temperature $T_e$ is computed from the caloric curve of the reactant Ind$^+$. 

\begin{figure}
\includegraphics[width=\columnwidth]{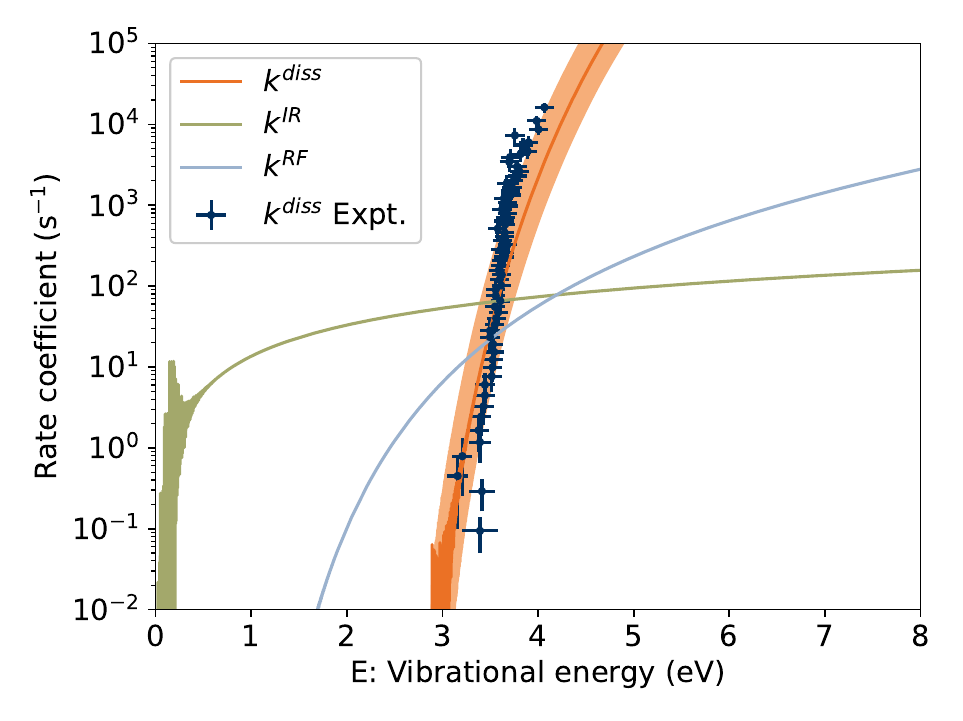}
\caption{Rate coefficients for Ind$^+$: dissociation by H-loss $k^{diss}(E)$ determined from KER distributions, IR vibrational radiative cooling $k^{IR}(E)$ from SHC model (Eq.~\ref{eq_kir}), and recurrent fluorescence $k^{RF}(E)$ from AIMD simulations (FIG.~\ref{fig_rf}).}
\label{fig_coeffs}
\end{figure}

From the fit shown in FIG.~\ref{fig_arrh}, we find the activation energy $E_a=2.9(1)$~eV and the pre-exponential factor $A=3.3(5)\times 10^{14}$~s$^{-1}$. Our experimentally determined microcanonical dissociation rate coefficient $k^{diss}(E)=k_m(t)$ is plotted in FIG.~\ref{fig_coeffs} as a function of the vibrational energy of the reactant, $E=E_{tot}(T_p)+E_a$, along with our calculated radiative cooling rate coefficients $k^{IR}(E)$ and $k^{RF}(E)$. 

Our activation energy agrees with the 2.6(4)~eV found by West \textit{et al.} for Ind$^+$ using a photo-electron photo-ion coincidence technique \cite{West2014}. A weighted average of threshold energies for $sp^3$-hybridized H atom losses from six different PAH cations determined by similar methods give $E_a=2.4(4)$~eV \cite{West2014,West2014a,West2018}. Our own quantum chemical calculations found dissociation energies for H-loss from Ind$^+$ of 2.8~eV at both the DSD-PBEP86-D3(BJ)/def2-TZVPP and CCSD(T)/cc-pVTZ levels of theory, in good agreement with our our measured value.

The pre-exponential factor is close to the `universal' Gspann value of 1.6$\times 10^{15}$~s$^{-1}$ for the evaporation of atomic and molecular clusters \cite{Klots1976}. In an RRKM framework, our value of $A$ corresponds to an activation entropy of $\Delta S^{\ddag}_{\mathrm{1000 K}}=14(1)$~JK$^{-1}$mol$^{-1}$, consistent with a loose transition state. West \textit{et al.} \cite{West2014} give $\Delta S^{\ddag}_{\mathrm{1000 K}}=-2\pm 38$~JK$^{-1}$mol$^{-1}$ for Ind$^+$, and an average value \cite{West2014,West2014a,West2018} for $sp^3$-hybridized H-losses of $\Delta S^{\ddag}_{\mathrm{1000 K}}=44\pm 20$~JK$^{-1}$mol$^{-1}$. 

Taking instead a detailed balance perspective, the cross section for H atom capture by the indenyl cation is given by \cite{Andersen2001}:
\begin{equation}
\sigma_c=\frac{\pi^2\hbar^3}{2\mu(k_BT_p)^2}A
\label{eq_xsec}
\end{equation}
where $\mu$ is the reduced mass. For the average product temperatures determined from the fit to the KER distributions (Eq.~\ref{eq_kerd}), Eq.~\ref{eq_xsec} gives $\sigma_c=1.2(2)$~\AA$^2$. This may be compared to the Langevin cross section, which gives an upper limit on the reaction rate \cite{Andersen2001}:
\begin{equation}
\sigma_{L}=\pi\sqrt{\frac{\pi\alpha q_e^2k_e}{2k_BT_d}}
\end{equation}
where $\alpha=0.667$~\AA$^3$ is the polarizability volume of the H atom fragment, $q_e$ is the elementary charge, and $k_e=1/4\pi\epsilon_0$. This yields $\sigma_L=49.2(1)$~\AA$^2$, which is comparable to the geometric cross section of indene \cite{Hashemi2023}. Thus, our measured dissociation rate for Ind$^+$ implies a formation rate more than an order of magnitude below the Langevin rate.

\subsection{Dissociation rate}
\label{sec_rate}

\begin{figure}
\includegraphics[width=\columnwidth]{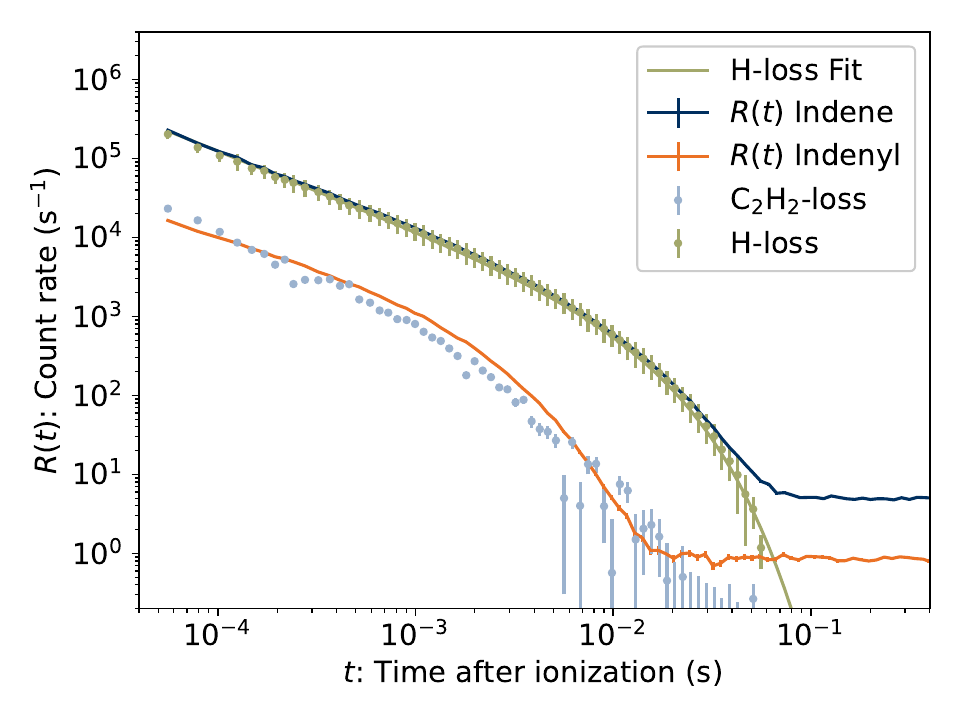}
\caption{Measured total count rate $R(t)$ of neutral fragments of Ind$^+$. Points are partial yields of H and C$_2$H$_2$ determined from fits of KER distributions. Scaled count rate for [Ind-H]$^+$, measured separately, is plotted for comparison to the C$_2$H$_2$ yield. We attribute the C$_2$H$_2$ contribution to $^{13}$C [Ind-H]$^+$ co-stored with isobaric Ind$^+$.}
\label{fig_spon}
\end{figure}

The total neutral product count rate $R(t)$ for Ind$^+$ ions stored in DESIREE is shown in FIG.~\ref{fig_spon} (blue curve). For the first few milliseconds after injection, the rate follows a power-law $R(t)\propto t^{-1}$, indicative of a broad internal energy distribution \cite{Hansen2001}. After about 10~ms, the rate begins to deviate from the power law trend as radiative cooling becomes competitive with dissociation. After 100~ms, a constant rate is observed due to dissociation of stored ions induced by collisions with the residual gas in the ring. 

The total count rate $R(t)$ is divided into constributions from H-loss from C$_9$H$_8^+$ (green symbols in FIG.~\ref{fig_spon}) and C$_2$H$_2$-loss from the contaminant $^{13}$C$^{12}$C$_8$H$_7^+$ (blue symbols) according to fits of the KER distribution as in FIG.~\ref{fig_lumps}. The count rate for a separate measurement of pure $^{12}$C indenyl (C$_9$H$_7^+$), scaled to that of the C$_2$H$_2$-loss contamination signal, is included for comparison (orange curve). It is clear that H-loss from Ind$^+$ is the dominant contribution to $R(t)$, and the C$_2$H$_2$-loss channel will therefore be neglected in the remaining analysis.  

The spontaneous decay rate $R(t)$ was fit with
\begin{equation}
R(t)=r_0t^{-1}e^{-k_ct}
\label{eq_rate}
\end{equation}
\noindent where the dimensionless constant $r_0$ collects all the experimental parameters and $k_c$ is the critical rate coefficient where cooling competes with dissociation. The value of $k_c$ according to the fit is 78(1)~s$^{-1}$, which is lower than the values obtained for other cationic, pure hydrocarbon PAHs to date, which range from 460(30)~s$^{-1}$ for naphthalene \cite{Lee2023} to over 1000~s$^{-1}$ for tetracene \cite{Bernard2023}. This can be explained by low dissociation threshold energy for loss of the labile $sp^3$ hybridized H atom. The fully $sp^2$-hybridized perylene cation C$_{20}$H$_{12}^+$, for example, decays mainly by H-loss but with a higher dissociation energy of 4.8(6)~eV \cite{West2018}, resulting in a value of $k_c=450(2)$~s$^{-1}$ \cite{Stockett2020b}. Critical rate coefficients on the order of 100~s$^{-1}$ have been determined for other functionalized PAH cations and explained by infrared radiative cooling \cite{Zhu2022}, with larger values being interpreted as evidence for RF \cite{Stockett2023}. 

\begin{figure}
\includegraphics[width=\columnwidth]{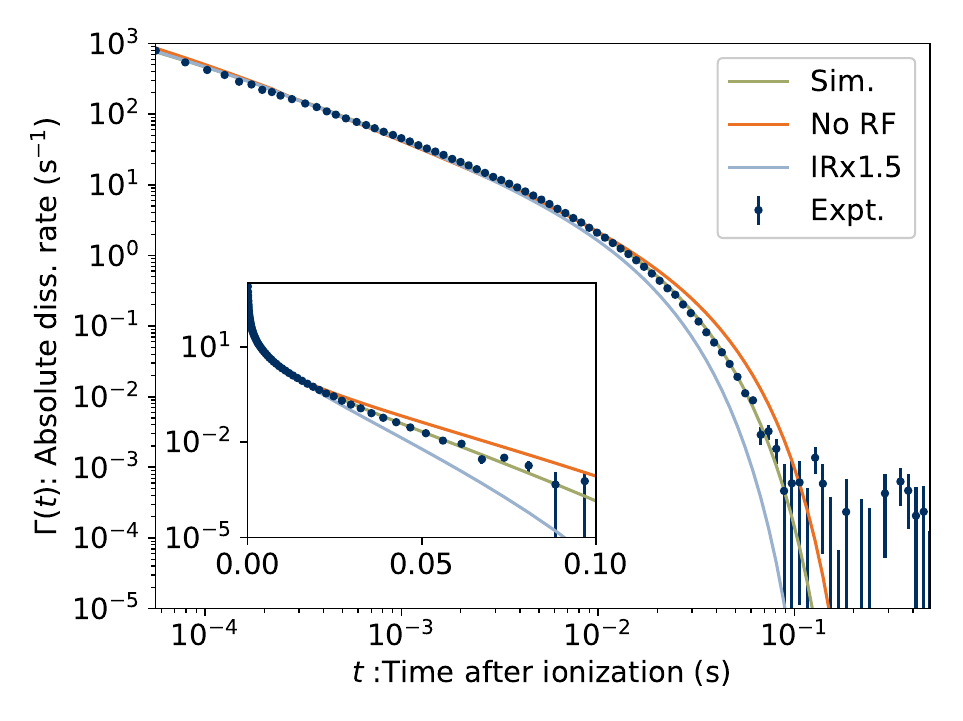}
\caption{Normalized spontaneous decay rate $\Gamma(t)$ (Eq.~\ref{eq_Gamma}) of Ind$^+$. The solid curves are the results of master equation simulations for three different assumptions about the radiative cooling rates. `Sim.': rates as calculated with full model described in this work; `No RF': $k^{RF}$ set to zero; `IR$\times 1.5$': $k^{RF}$ as calculated, $k^{IR}$ increaed by a factor of 1.5. Inset with linear time axis.}
\label{fig_gamma}
\end{figure}

The absolute, per-particle dissociation rate $\Gamma(t)$ is related to the measured count rate $R(t)$ by:
\begin{equation}
\Gamma(t)=\frac{f_{rev}q_e}{\eta_{det}GI_{avg}}R(t),
\label{eq_Gamma}
\end{equation}
\noindent where $f_{rev}=43.07$~kHz is the revolution frequency of the ions in the storage ring, $q_e$ is the elementary charge, $\eta_{det}$ is the detector efficiency, $G=0.11$ is the fraction of the circumference of storage ring visible to the detector (see FIG.~\ref{fig_desiree}) and $I_{avg}=160(1)$~pA is the average ion beam current measured at the end of each storage cycle.

Results of our master equation simulations compared to the the measured dissociation rate $\Gamma(t)$ are presented in FIG.~\ref{fig_gamma}. Simulations were run for a range of temperatures $T_0$ specifying the initial, Boltzmannian distribution of vibrational energies of the ensemble. The detector efficiency $\eta_{det}$ was also taken as a free parameter. The simulation that best reproduced the measured data had $T_0=1650(40)$~K and $\eta_{det}=0.113(1)$. This temperature is comparable to that obtained for other PAHs using the same experimental and modeling techniques, and the derived detector efficiency is reasonable for hydrogen atoms with $\approx 740$~eV kinetic energy in the lab frame as in the present experiment \cite{Barat2000,Stockett2020b}.

Two alternative simulation results are also presented in FIG.~\ref{fig_gamma}. In the first, to test our hypothesis that vibrational motion enhances the RF rate, we turn off RF in the simulation. The best fitting simulation under these conditions ($T=1800(100)~K$ and $\eta_{det}=0.101(2)$) diverges from the measured rate after a few tens of ms. This shows that RF plays a role in the simulated cooling dynamics despite the RF rate coefficient being lower than for IR cooling in the energy window probed in our experiment. A second scenario leaves RF in place but increases the IR cooling rate by a factor of 1.5. Differences in the IR rates on this scale are sometimes noted between different levels of theory \cite{Bull2019a}. Again, the best fitting simulation ($T=1670(40)~K$ and $\eta_{det}=0.108(1)$) diverges from the measured data, but in the opposite direction. Thus, an overestimate of the RF rate coefficient could be compensated by an underestimate of the IR rate coefficient. We conclude that RF makes, at most, a small contribution to the radiative stabilization of indene cations.

\subsection{On the connection between measured rates and rate coefficients}

In our analysis of the dissociation rate, we make the assumption that the measured decay rate $R(t)$ is directly proportional to dissociation rate coefficient of the ions dissociating at time $t$ (Eq.~\ref{eq_km}), and that the constant of proportionality $r_0$ is determined from the fit of Eq.~\ref{eq_rate} to $R(t)$. This is motivated by the fact that $k^{diss}(E)$ increases rapidly with energy, so that the ensemble averaged decay rate is dominated by ions with energies $E_m(t)$ at the high-energy front of the internal energy distribution \textit{i.e.} $R(t)\propto k_m(t)\equiv k^{diss}(E_m(t))$, where we take the values of $E_m(t)$ to be those obtained from the KER distributions as described in Sec.~\ref{sec_kdiss}. We name the fraction of ions with $E\approx E_m(t)$ to be $\gamma_0(t)$, which can also be written as the ensemble averaged destruction probability:
\begin{equation}
\gamma_0(t)=\int g(E,t)(1-e^{-k^{diss}(E)t})dE/\int g(E,t)dE.
\end{equation}
Here, we take $\gamma_0$ to be constant and define
\begin{equation}
r_0=\gamma_0\frac{\eta_{det}GI_{avg}}{f_{rev}q_e}.
\label{eq_r0}
\end{equation}
From Eqs.~\ref{eq_km}, \ref{eq_Gamma}, and \ref{eq_r0}, $\Gamma(t)=\gamma_0k_m(t)$. The assumption that $\gamma_0$ is constant is equivalent to assuming that the number of stored ions is constant and equal to the ion beam current $I_{avg}$ measured at the end of the storage cycle. This condition is typically fulfilled in storage ring experiments, where only a few per cent of the ions in the stored ensemble have sufficient energy to decay prior to the onset of radiative cooling. 

\begin{figure}
\includegraphics[width=\columnwidth]{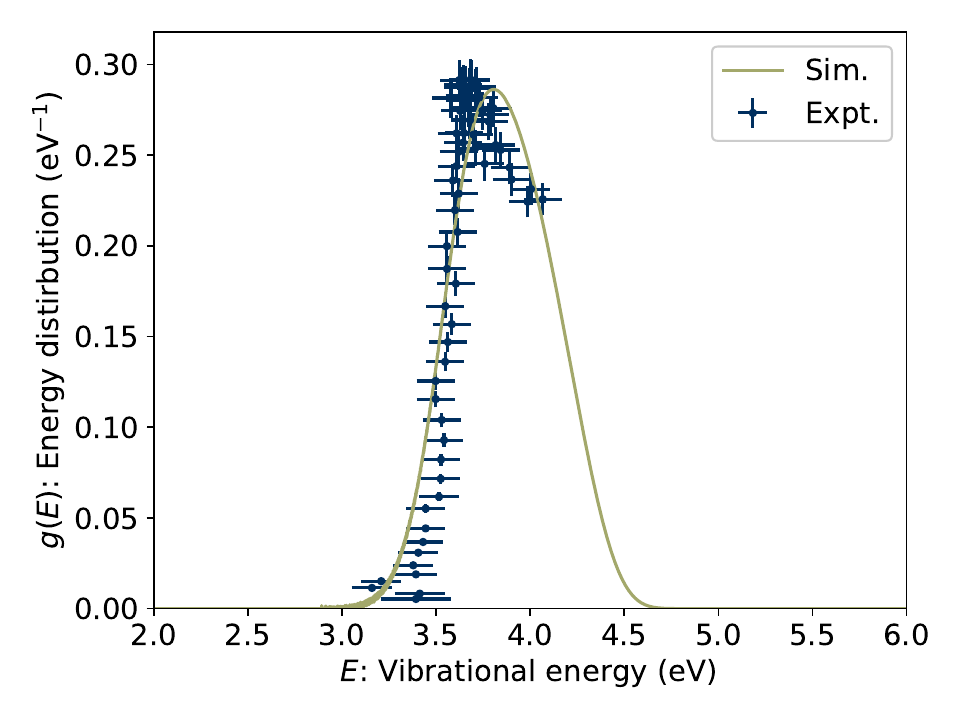}
\caption{Internal energy distribution $g^{\dagger}(E)$ of dissociating Ind$^+$ from Eq.~\ref{eq_klavs} (points) and from our master equation modeling (solid curve).}
\label{fig_g0}
\end{figure}

In the context of thermal electron emission from fullerene anions, Hansen \cite{Hansen2020} developed a model including the following expression for $\gamma_0$ (recast in our symbols):
\begin{equation}
\Gamma(t)=\gamma_0k_m(t)=g(E_m(t),t)\frac{E_aC}{t[\ln(At)]^2},
\label{eq_klavs}
\end{equation} 
where $g(E_m(t),t)$ and the heat capacity $C$ are assumed to be constant, and the most-likely rate coefficient is given by $k_m(t)=t^{-1}$. Solving instead for $g(E_m(t),t)$, and using our values for $E_m(t)$ obtained from the KER distributions, and our Arrhenius parameters $A$ and $E_a$ determined from the fit of Eq.~\ref{eq_arrh}, we obtain the vibrational energy distribution of ions which dissociated during the experiment, denoted $g^{\dagger}(E)$ \cite{Zhu2022}, plotted in FIG.~\ref{fig_g0}. Here we assume a constant heat capacity $C=45/k_B$ and again take the detector efficiency as a free parameter to obtain agreement with the $g^{\dagger}(E)$ distribution computed in our best-fitting master equation simulation. We find $\eta_{det}=0.095(2)$, which is similar to the values found comparing the simulated and measured dissociation rates (Sec.~\ref{sec_rate}). The qualitative agreement in the shape of the distributions, and the quantitative agreement of their magnitudes, validates our assumption in Eq.~\ref{eq_km} that the measured dissociation rate $R(t)$ can be treated as being directly proportional to the dissociation rate coefficient $k_m(t)$.

\subsection{Radiative cooling}

\begin{figure}
\includegraphics[width=\columnwidth]{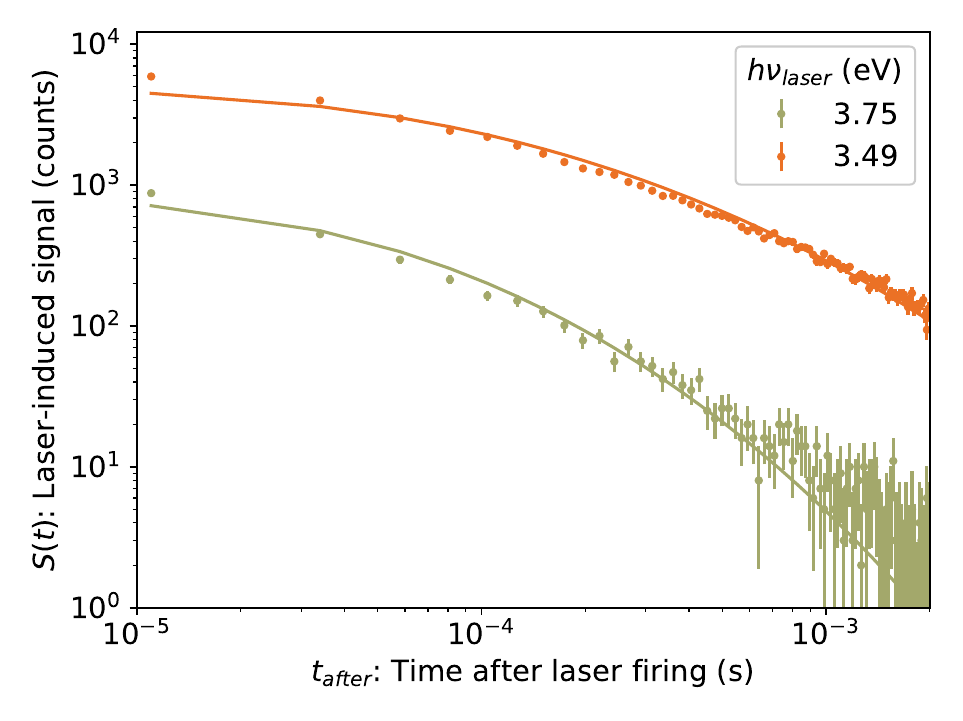}
\caption{Laser-induced neutral fragment yield $S(t_{after})$ of Ind$^+$ for two different laser excitation energies at the same laser firing time $t=0.6$~s.}
\label{fig_lasercounts}
\end{figure}

Only a fraction of the ions injected into the DESIREE storage ring decay by the dissociation process discussed so far. Most will continue to circulate while cooling by emission of thermal radiation. The evolution of the internal energy distribution of the stored ions $g(E,t)$ can be tracked with laser probing techniques \cite{Sunden2009,Stockett2020b,Rasmussen2022,NavarroNavarrete2023}. FIG.~\ref{fig_lasercounts} shows the yield of neutral fragments $S$ as a function of time following laser excitation $t_{after}$ with $h\nu_{laser}=3.75$~eV (330~nm) and 3.49~eV (355~nm) photons, for a laser firing time of $t=0.6$~s. 

To analyze the laser-induced dissociation rates, we fit a simplified model:
\begin{equation}
S(t_{after})=s_0\int k^{diss}(E)g_{T}(E)e^{-k^{tot}t_{after}}dE,
\label{eq_simp}
\end{equation}
where $s_0$ includes the absorption probability and the various experimental parameters, $g_{T}$ is a Boltzmann distribution with the temperature $T$, shifted by the laser photon energy $h\nu_{laser}$ and $k^{tot}=k^{diss}+k^{IR}+k^{RF}$. The solid lines in FIG.~\ref{fig_lasercounts} are fits of Eq.~\ref{eq_simp}. In contrast to our master equation simulations detailed above, this model neglects the re-distributive effect of radiative cooling. Essentially, it assumes that a single photon is always sufficient to quench dissociation, and may thus underestimate the dissociation rate for the hottest ions. This would not appear to be a serious issue. A fit of Eq.~\ref{eq_simp} to the spontaneous dissociation rate gave $T=1695(5)$~K, which is consistent with the more accurate master equation approach given in Sec.~\ref{sec_rate}. 

\begin{figure}
\includegraphics[width=\columnwidth]{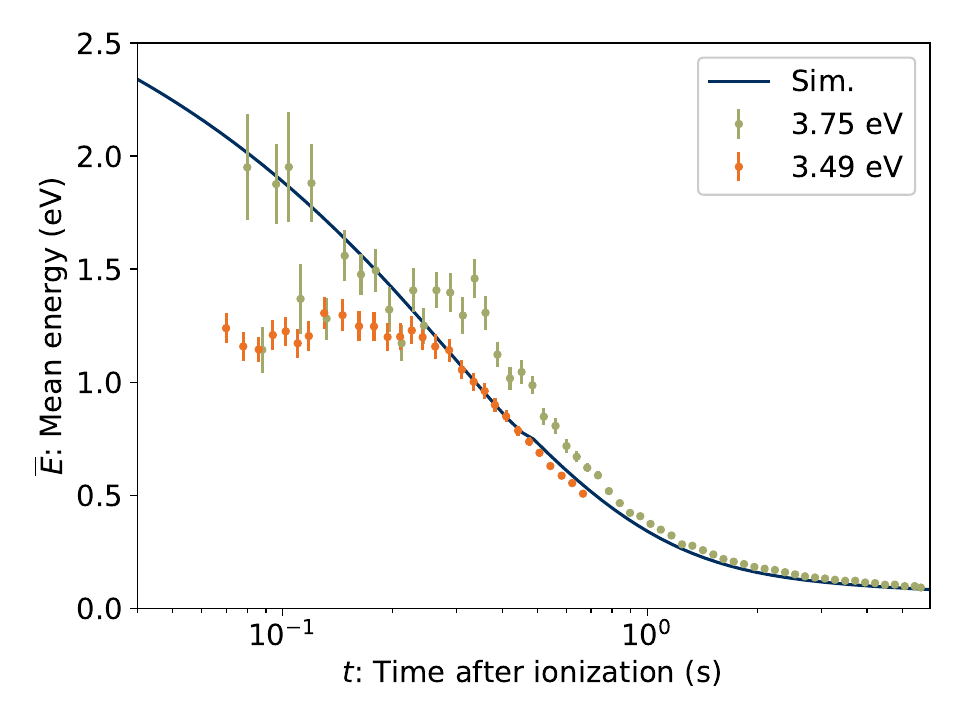}
\caption{Mean energies of the Boltzmann model (Eq.~\ref{eq_simp}) fitted to the measured $S(t_{after})$ curves for Ind$^+$ at two different laser photon energies. The solid line is the mean energy of the ensemble in our best-fitting master equation simulation (see Sec.~\ref{sec_rate}).}
\label{fig_emean}
\end{figure}

The mean energies $\overline{E}$ of the Boltzmann distributions fit to the laser-induced decay curves are plotted in FIG.~\ref{fig_emean}. Good agreement is generally found between these fitted energies and the mean energy of the ensemble in our master equation simulation which best fits the spontaneous dissociation rate (solid curve in FIG.~\ref{fig_emean}).

\begin{figure}
\includegraphics[width=\columnwidth]{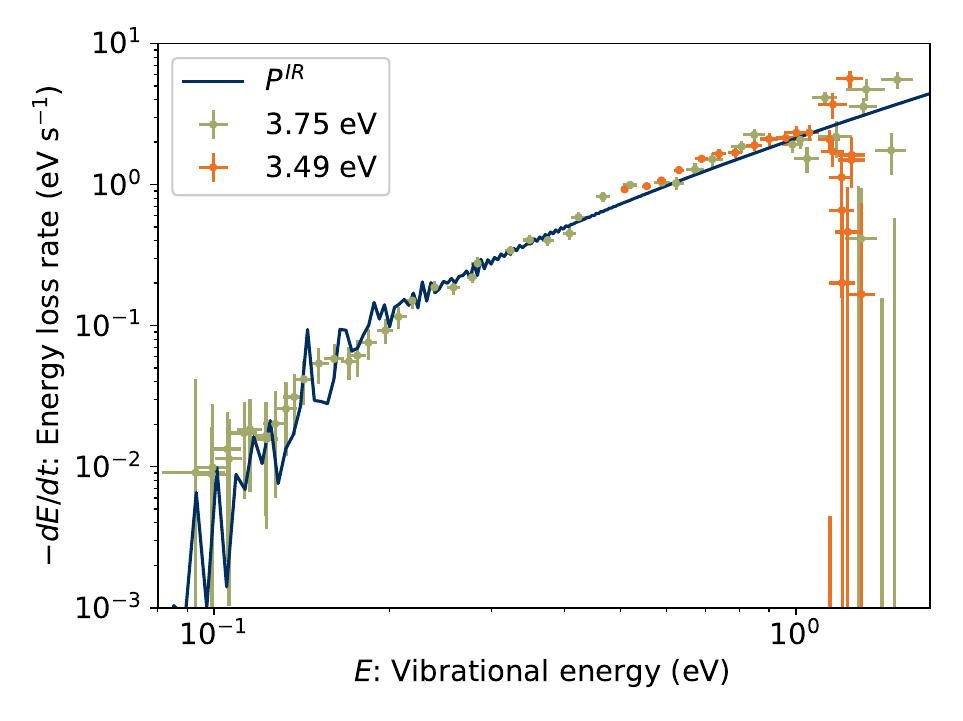}
\caption{Energy loss rate for Ind$^+$ computed as the gradient of the fitted $\overline{E}$ values (FIG.~\ref{fig_emean}). The solid curve is the vibrational radiative power.}
\label{fig_pow}
\end{figure}

The energy loss rate due to radiative cooling is computed as the time derivative of $\overline{E}$, plotted in FIG.~\ref{fig_pow}. Good agreement is found with the calculated total vibrational radiative power $P^{IR}=\sum_sK^{IR}_sh\nu_s$, supporting the accuracy of our calculated rate coefficients which, as discussed in Sec~\ref{sec_rate}, imply a minor role for RF in the radiative stabilization of Ind$^+$.

\subsection{Astrochemical implications}

\begin{figure}
\includegraphics[width=\columnwidth]{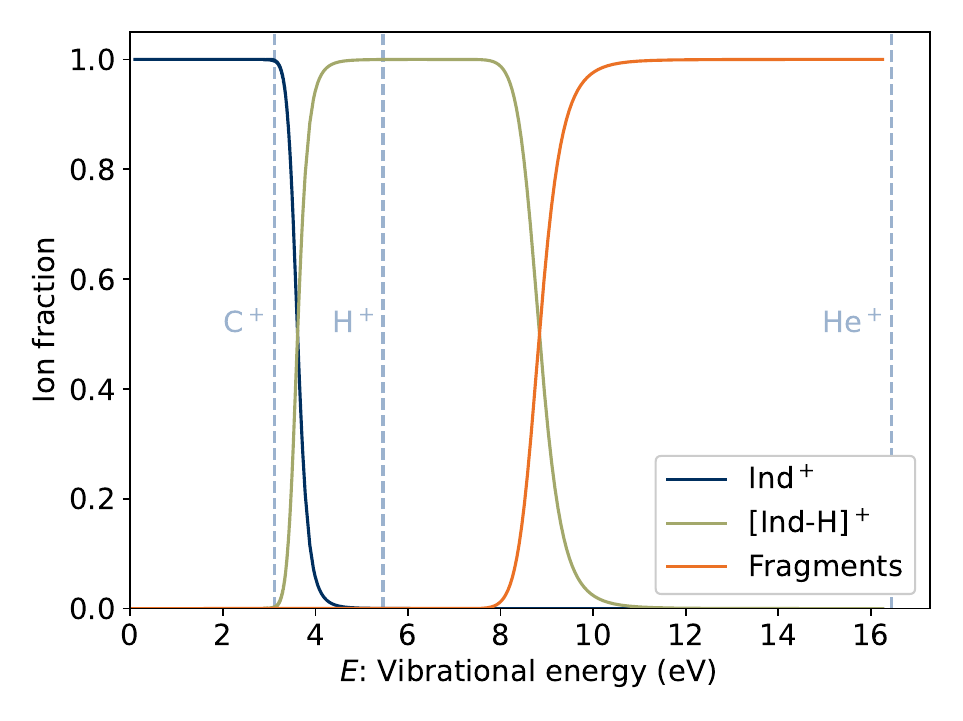}
\caption{Breakdown diagram for products of ionizing collisions between neutral Ind and atomic cations included in astrochemical models. The dashed vertical lines indicate the maximum energy transferred to Ind$^+$ for each atomic collision partner.}
\label{fig_breakdown}
\end{figure}

Astrochemical models routinely underestimate the observed abundances of small PAHs in TMC-1 \cite{McGuire2021,Burkhardt2021}. The main destruction mechanism included in these models is collisions with the cations H$^+$, He$^+$, C$^+$, H$_3^+$ and HCO$^+$, which are assumed to lead inexorably to complete degradation of the PAHs into small linear fragments \cite{McGuire2021}. These models may be improved by including more complete product breakdown reactions. In FIG.~\ref{fig_breakdown}, we give a breakdown diagram for indene following ionizing collisions based on the dissociation and radiative cooling rate coefficients for Ind$^+$ (FIG.~\ref{fig_coeffs}) and [Ind-H]$^+$ from Ref.~\cite{Bull2025}. At the low temperatures ($\sim$10~K) prevalent in TMC-1, the atomic cations will undergo electron transfer at large distances  with negligible translational-to-vibrational energy transfer \cite{Bohme1992}, leaving the ionized PAH with a maximum internal excitation energy equal to the difference in the ionization potentials (the molecular cations will lead to proton transfer). Note that these maximal energy transfers, given as vertical dashed lines in FIG.~\ref{fig_breakdown}, are extremely unlikely in distant electron-transfer reactions. With its relatively low ionization potential (11.26 eV), charge transfer forming carbon atoms occurs at shorter distances and will lead to greater heating of the Ind$^+$ product. Collisions with C$^+$ are thus considered the most important PAH-destruction channels in many astrochemical models \cite{Canosa1995,LePage2001}. However, given the 8.14~eV ionization potential of Ind \cite{West2014}, such collisions yield a maximal energy transfer of 3.12~eV. This is unlikely to induce dissociation of Ind$^+$, which is stabilized by radiative cooling up to a critical energy of 3.6~eV, where the dissociation and radiative rate coefficients are equal. Charge exchange reactions with protons could transfer sufficient energy to induce H-loss from Ind$^+$, yielding [Ind-H]$^+$ though, as stated above, such high high-energy transfers are rather unlikey in distant electron-transfer reactions. Collision with He$^+$ could potentially transfer sufficient energy to further break down [Ind-H]$^+$. However, the second ionization potential of Ind, at 21.8(1)~eV \cite{Roithova2006}, is low enough that formation of Ind$^{2+}$ may be competitive. 

It seems clear that collisions between PAHs and atomic cations in molecular clouds will primarily produce radiatively stabilized PAH cations. Previous studies have found that these cations do not react with H$_2$, and react mainly by association with H, N, and O atoms to form larger cationic molecules \cite{Snow1998,Betts2006}. Destruction of PAH cations in dark clouds is expected mainly in recombination reactions with free electrons \cite{Wakelam2008} or mutual neutralization with anions including PAH anions \cite{Lepp1988}. However, given that neutral PAHs may also be radiatively stabilized by RF \cite{Bull2025}, it is not obvious that extensive fragmentation should be expected following recombination.

\section{Conclusion}

Through analysis of the kinetic energy released upon H atom loss from indene cations, we determine a dissociation rate coefficient consistent with results of an earlier photodissociation study \cite{West2014}. The shape of the KER distribution is consistent with a qualitative model of tunneling through an intersection of potential energy surfaces \cite{Gridelet2006,Hansen2018}. More experiments on H-loss reactions for other hydrocarbons are needed to asses how general this phenomenon might be.

Using our best estimates of the vibrational and electronic radiative cooling rate coefficients, combined with the dissociation rate coefficient deduced from the present experiments, we find a limited role for recurrent fluorescence in the stabilization of indene cations. This is noteworthy as RF has been found to be the dominant cooling mechanism of nearly all PAH cations investigated experimentally to date \cite{Martin2013,Martin2015,Ji2017,Saito2020,Stockett2020b,Stockett2023,Lee2023,NavarroNavarrete2023,Bernard2023}. Still, we conclude that the IR radiative cooling rate is sufficiently high to prevent dissociation following ionizing collisions with carbon cations. This may partly explain the high observed abundance of indene in interstellar clouds \cite{Cernicharo2021}, demonstrating the importance of radiative stabilization of small PAHs in astrochemical reaction networks.

\section*{Acknowledgments}

This work was supported by Swedish Research Council grant numbers 2023-03833 (HC), 2020-03437 (HZ), Knut and Alice Wallenberg Foundation grant number 2018.0028 (HC and HZ), EPSRC grant EP/W018691 (JNB), Olle Engkvist Foundation grant number 200-575 (MHS), and Swedish Foundation for International Collaboration in Research and Higher Education (STINT) grant number PT2017-7328 (JNB and MHS). We acknowledge the DESIREE infrastructure for provisioning of facilities and experimental support, and thank the operators and technical staff for their invaluable assistance. The DESIREE infrastructure receives funding from the Swedish Research Council under the grant numbers 2021-00155 and 2023-00170.

\section*{Data availability}

The data that support the findings of this study are available from the corresponding author upon reasonable request.

\section*{Conflict of interest}

The authors have no conflicts to disclose.

\section*{Appendix}

Here we describe the implementation of matrix transforms used to compute:
\begin{enumerate}
\item the total density distribution of the Newton sphere consisting of contributions from neutral particles emitted with a given kinetic energy distribution along a linear segment normal to the image plane, and
\item the forward Abel transform between the Newton sphere density distribution and the radial intensity distribution in the image plane. 
\end{enumerate}

These matrix transforms are used to fit model kinetic energy release (KER) distributions $P(\epsilon)$ to measured radial intensity distributions $P(r)$ measured with a two-dimensional microchannel plate detector after linear translation of the expanding Newton sphere from the point of reaction to the detector plane. This approach is preferred as the measured $P(r)$ distributions have well-defined uncertainties given by counting statistics, while inverting the transforms inevitably introduces new errors.

\subsection*{A1: Spatial transform}

For a point source, there is a one-to-one relationship between the velocity of an emitted particle and the radius of the Newton sphere at a given time after emission. In particle imaging experiments, where the lab-frame velocity of the reactant ion is much higher then the velocity of the fragment in the center-of-mass frame, the relevant time after emission corresponds to a translation of the center of the Newton sphere to the detector plane, and the kinetic energy of the particle maps directly to spherical radius $\rho$ as given by:
\begin{equation}
\epsilon(\rho,L) = \frac{m_{neut}}{m_{cat}}E_{acc}\left(\frac{\rho}{L}\right)^2
\label{eq_epsilonL}
\end{equation}
where $L$ is the distance from the point of decay to the detector. The Newton sphere density $P(\rho)=\frac{\partial k^{diss}}{\partial \rho}$ may be obtained from the KER distribution $P(\epsilon)=\frac{\partial k^{diss}}{\partial \epsilon}$ by:
\begin{equation}
P(\rho)=\frac{\partial \epsilon}{\partial \rho}P(\epsilon)=\frac{2\epsilon}{\rho}P(\epsilon)
\label{eq_epstorho}
\end{equation}
where Eq.~\ref{eq_epsilonL} was used.

In DESIREE, particles are emitted along a linear segment with a length $L_{SS}=0.96$~m, which is comparable to the distance $L_{mid}=1.7$~m from the midpoint of this segment to the detector. Thus, the density at a radius $\rho_i$ ($i=0,1\ldots n$, where $n$ is the radius of the detector in pixels) of the cumulative Newton sphere includes contributions from energies ranging from $\epsilon_{j_{min}}(\rho_i)=\epsilon(\rho_i,L_{max})$ to $\epsilon_{j_{max}}(\rho_i)=\epsilon(\rho_i,L_{min})$, where $L_{max}=L_{mid}+L_{SS}/2$ and $L_{min}=L_{mid}-L_{SS}/2$. Eq.~\ref{eq_epstorho} thus becomes the weighted average:
\begin{equation}
P(\rho_i)=\sum_{j=j_{min}}^{j_{max}}\mathcal{T}_{ij}P(\epsilon_j)\,
\end{equation}
where
\begin{equation}
\mathcal{T}_{ij}=\frac{2\epsilon_j}{\rho_i(j_{max}-j_{min})}.
\end{equation}

We define our energy coordinate vector $\epsilon_j$ ($j=0,1\ldots n$) to span the range from $\epsilon_{j_{min}}(\rho_0)$ to $\epsilon_{j_{max}}(\rho_n)$. We choose the $\epsilon_i$ to increase quadratically, to maintain the scaling of Eq.~\ref{eq_epsilonL}.

\begin{figure}
\includegraphics[width=\columnwidth]{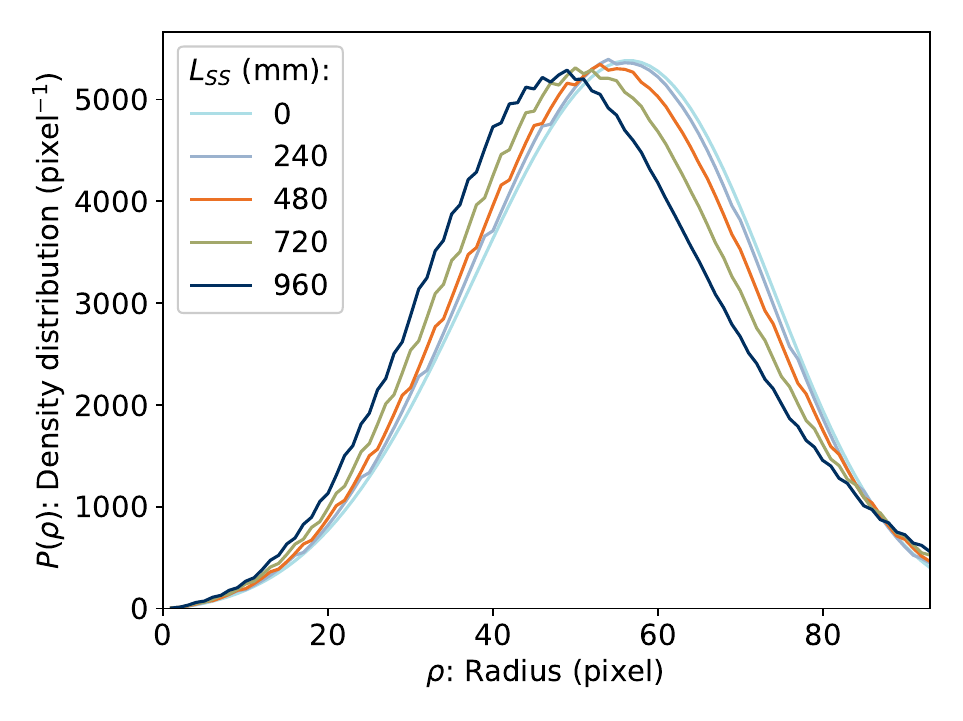}
\caption{Transformations of a model KER distribution to integrated Newton sphere density for varying lengths of the emitting segment $L_{SS}$.}
\label{fig_rho}
\end{figure}

In FIG.~\ref{fig_rho}, we show the results of several transformations of a model KER distribution into integrated Newton sphere density distributions using $\mathcal{T}$ matrices with varying lengths of the emitting segment $L_{SS}$. The model distribution is of the form of Eq.~\ref{eq_kerd} using the parameters found from the fit to the late-time KER distribution of Ind$^+$ (FIG.~\ref{fig_kerd}). The actual value of $L_{SS}$ in DESIREE is 960~mm, leading to a density distribution which is notably distorted relative to that from a point source.

\begin{figure}
\includegraphics[width=\columnwidth]{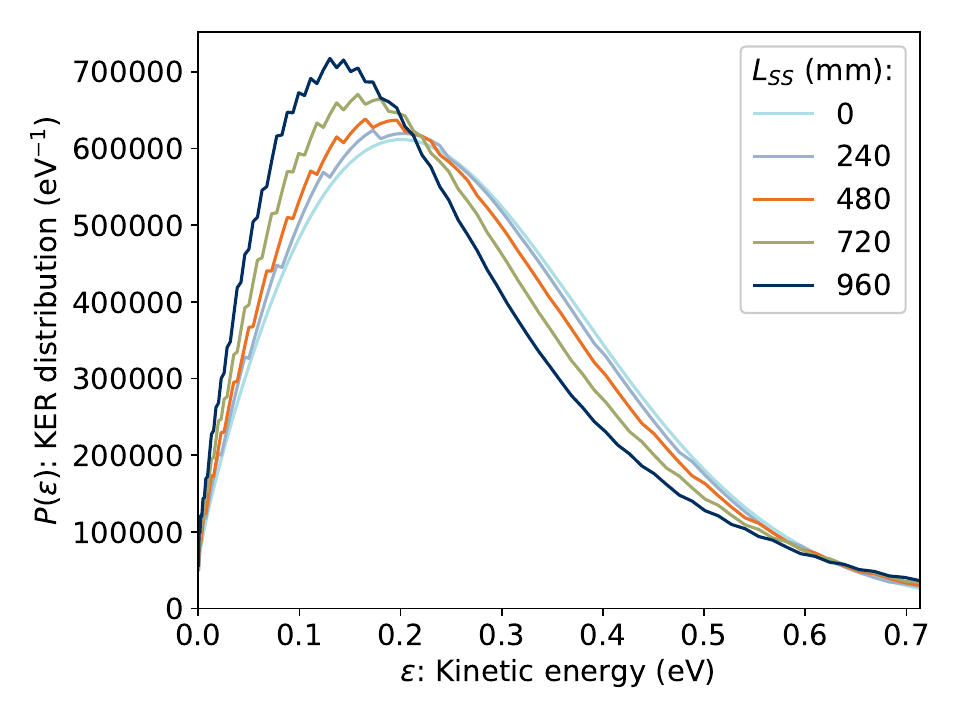}
\caption{Pseudo-calibrations of transformed model KER distribution for varying lengths of the emitting segment $L_{SS}$.}
\label{fig_cal}
\end{figure}

Unfortunately, the $\mathcal{T}$ matrix is ill-conditioned and cannot be inverted. In order to plot KER distributions, it is thus necessary to perform a pseudo-calibration by Eq.~\ref{eq_epsilonL} with $L=L_{mid}$. As shown in FIG.~\ref{fig_cal}, this results in distorted KER distributions for emitting segments of finite length. We reiterate that all numerical results are obtained by fitting of the radial intensity in the detector plane which, because it only requires the forward transform, is not affected by this distortion.

\subsection*{A2: Abel transform}

The Abel transform of a one-dimensional distribution $f(\rho)$ is given by:
\begin{equation}
F(r)=\int_{\rho}^\infty\frac{2\rho f(\rho)}{\sqrt{\rho^2-r^2}}dr.
\end{equation}
We operate on the Newton sphere density distribution integrated over solid angle, $P(\rho)=4\pi \rho^2f(\rho)$, and thus desire a matrix approximating the integral: 
\begin{equation}
F(r)=\int_y^\infty\frac{P(\rho)}{2\pi \rho\sqrt{\rho^2-r^2}}dr.
\end{equation}
We may naively take this matrix to be:
\begin{equation}
\mathcal{A}_{ij} = \frac{1}{2\pi\rho_i\sqrt{\rho_i^2-r_j^2}} \qquad (i\geq j),
\end{equation}
and $\mathcal{A}_{ij}=0$ for $i<j$, for two coordinate vectors $\rho_i=r_i=i$ for $i=0,1,\ldots (n_{pix}-1)$ where $n_{pix}$ is the width of the detector in pixels. This is problematic near the main diagonal, so instead we define the coordinate vectors so:
\begin{equation}
r_i=\rho_i-\delta \qquad (0<\delta<1),
\end{equation}
with $\delta$ being an adjustable parameter. Still we have a problem with $\mathcal{A}_{00}$, which we take to be a second adjustable parameter. The numerical value of these parameters is found by least squares fitting, either to analytical functions with known Abel transforms, or to real data using the two-dimensional Abel transform algorithms included in the PyAbel package \cite{pyabel}. We find that $\delta=0.35$ and $\mathcal{A}_{00}=0.8$ generally produce good results. Finally, to obtain the azimuthally integrated radial intensity on the detector, $P(r)$, we rescale $F(r)$ by the diagonal matrix:
\begin{equation}
\mathcal{R}_{ii}=2\pi r_i.
\end{equation}

\begin{figure}
\includegraphics[width=\columnwidth]{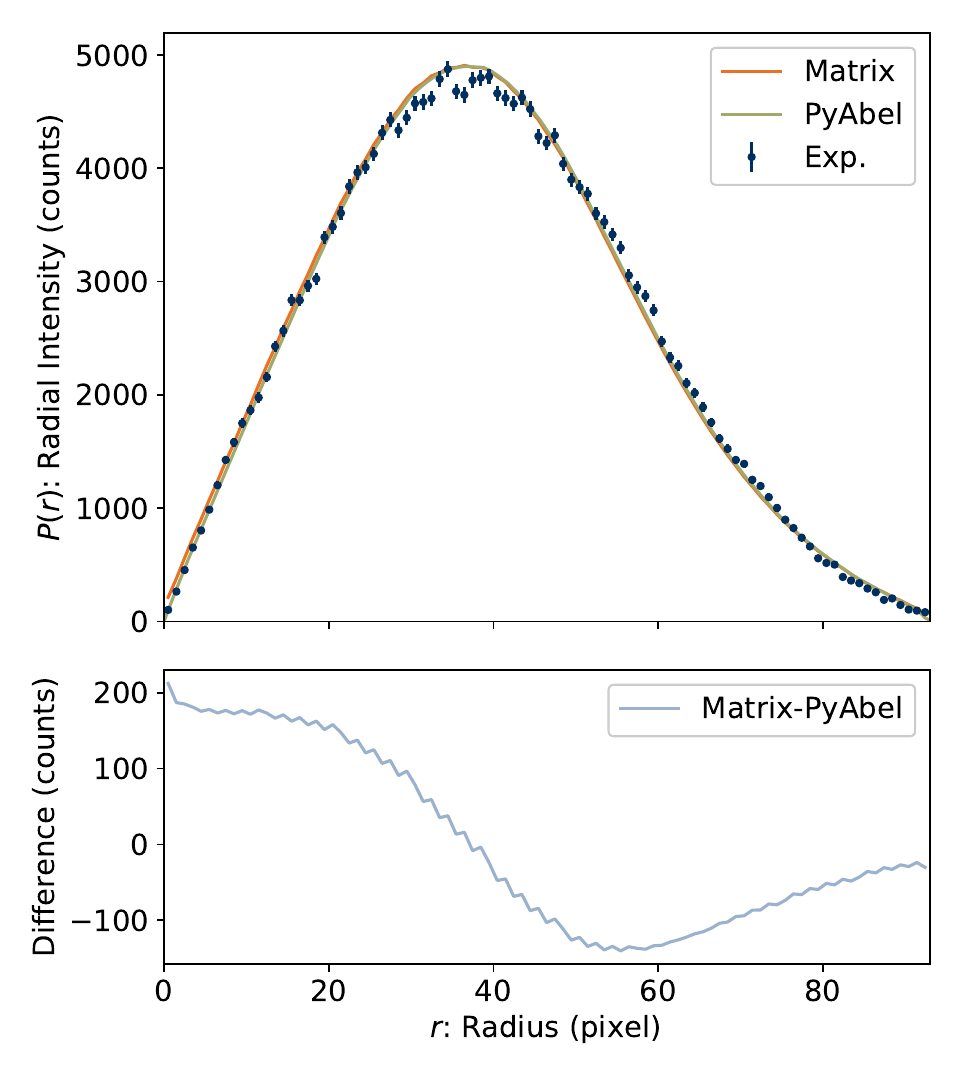}
\caption{Comparison of forward Abel transform methods for the model KER distribution, and the measured radial intensity distribution of Ind$^+$.}
\label{fig_rtot}
\end{figure}

In FIG.~\ref{fig_rtot}, we show a comparison of forward Abel transforms using the complete matrix transform $\mathcal{RA}$ and the PyAbel package. Both start from the model distribution $P(\rho)$ from FIG.~\ref{fig_rho} (for $L_{SS}=960$~mm). For PyAbel the Hansen-Law algorithm was employed, with other options giving nearly identical results. The differences in the resulting $P(r)$ distributions are smaller than the experimental uncertainties.  

\begin{figure}
\includegraphics[width=\columnwidth]{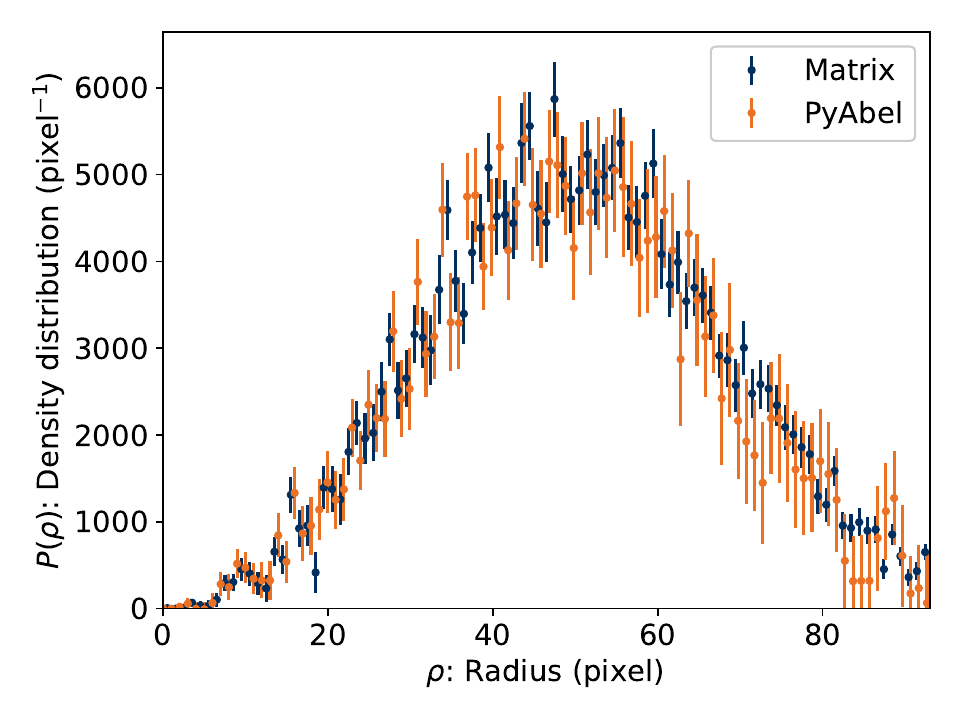}
\caption{Comparison of reverse Abel transform methods for the measured radial intensity distribution of Ind$^+$.}
\label{fig_invabel}
\end{figure}

With the given choices of $\delta$ and $\mathcal{A}_{00}$, the transform matrix is readily inverted using the standard Moore-Penrose algorithm. By reduction of dimensionality and cancellation of counting uncertainties by azimuthal integration prior to inversion, the matrix implementation of the reverse Abel transform outperforms the benchmark PyAbel routine. As described in Sec.~\ref{sec_exp}, the uncertainty in the density distribution $P(\rho)$ is found by inverting and averaging 128 $P(r)$ distributions with intensities normally distributed according to the measured distribution and its counting statistics. Applying this same approach using PyAbel, which operates on the full 2D image rather than the radial intensity distribution, takes 60 times longer than the matrix method. As shown in FIG.~\ref{fig_invabel}, the PyAbel result is also noisier.


\begin{thebibliography}{58}%
\makeatletter
\providecommand \@ifxundefined [1]{%
 \@ifx{#1\undefined}
}%
\providecommand \@ifnum [1]{%
 \ifnum #1\expandafter \@firstoftwo
 \else \expandafter \@secondoftwo
 \fi
}%
\providecommand \@ifx [1]{%
 \ifx #1\expandafter \@firstoftwo
 \else \expandafter \@secondoftwo
 \fi
}%
\providecommand \natexlab [1]{#1}%
\providecommand \enquote  [1]{``#1''}%
\providecommand \bibnamefont  [1]{#1}%
\providecommand \bibfnamefont [1]{#1}%
\providecommand \citenamefont [1]{#1}%
\providecommand \href@noop [0]{\@secondoftwo}%
\providecommand \href [0]{\begingroup \@sanitize@url \@href}%
\providecommand \@href[1]{\@@startlink{#1}\@@href}%
\providecommand \@@href[1]{\endgroup#1\@@endlink}%
\providecommand \@sanitize@url [0]{\catcode `\\12\catcode `\$12\catcode
  `\&12\catcode `\#12\catcode `\^12\catcode `\_12\catcode `\%12\relax}%
\providecommand \@@startlink[1]{}%
\providecommand \@@endlink[0]{}%
\providecommand \url  [0]{\begingroup\@sanitize@url \@url }%
\providecommand \@url [1]{\endgroup\@href {#1}{\urlprefix }}%
\providecommand \urlprefix  [0]{URL }%
\providecommand \Eprint [0]{\href }%
\providecommand \doibase [0]{http://dx.doi.org/}%
\providecommand \selectlanguage [0]{\@gobble}%
\providecommand \bibinfo  [0]{\@secondoftwo}%
\providecommand \bibfield  [0]{\@secondoftwo}%
\providecommand \translation [1]{[#1]}%
\providecommand \BibitemOpen [0]{}%
\providecommand \bibitemStop [0]{}%
\providecommand \bibitemNoStop [0]{.\EOS\space}%
\providecommand \EOS [0]{\spacefactor3000\relax}%
\providecommand \BibitemShut  [1]{\csname bibitem#1\endcsname}%
\let\auto@bib@innerbib\@empty
\bibitem [{\citenamefont {Tielens}(2008)}]{Tielens2008}%
  \BibitemOpen
  \bibfield  {author} {\bibinfo {author} {\bibfnamefont {A.~G. G.~M.}\
  \bibnamefont {Tielens}},\ }\bibfield  {title} {\enquote {\bibinfo {title}
  {Interstellar polycyclic aromatic hydrocarbon molecules},}\ }\href@noop {}
  {\bibfield  {journal} {\bibinfo  {journal} {Annu. Rev. Astron. Astrophys.}\
  }\textbf {\bibinfo {volume} {46}},\ \bibinfo {pages} {289--337} (\bibinfo
  {year} {2008})}\BibitemShut {NoStop}%
\bibitem [{\citenamefont {Burkhardt}\ \emph {et~al.}(2021)\citenamefont
  {Burkhardt}, \citenamefont {Long Kelvin~Lee}, \citenamefont {Bryan~Changala},
  \citenamefont {Shingledecker}, \citenamefont {Cooke}, \citenamefont {Loomis},
  \citenamefont {Wei}, \citenamefont {Charnley}, \citenamefont {Herbst},
  \citenamefont {McCarthy},\ and\ \citenamefont {McGuire}}]{Burkhardt2021}%
  \BibitemOpen
  \bibfield  {author} {\bibinfo {author} {\bibfnamefont {A.~M.}\ \bibnamefont
  {Burkhardt}}, \bibinfo {author} {\bibfnamefont {K.}~\bibnamefont {Long
  Kelvin~Lee}}, \bibinfo {author} {\bibfnamefont {P.}~\bibnamefont
  {Bryan~Changala}}, \bibinfo {author} {\bibfnamefont {C.~N.}\ \bibnamefont
  {Shingledecker}}, \bibinfo {author} {\bibfnamefont {I.~R.}\ \bibnamefont
  {Cooke}}, \bibinfo {author} {\bibfnamefont {R.~A.}\ \bibnamefont {Loomis}},
  \bibinfo {author} {\bibfnamefont {H.}~\bibnamefont {Wei}}, \bibinfo {author}
  {\bibfnamefont {S.~B.}\ \bibnamefont {Charnley}}, \bibinfo {author}
  {\bibfnamefont {E.}~\bibnamefont {Herbst}}, \bibinfo {author} {\bibfnamefont
  {M.~C.}\ \bibnamefont {McCarthy}}, \ and\ \bibinfo {author} {\bibfnamefont
  {B.~A.}\ \bibnamefont {McGuire}},\ }\bibfield  {title} {\enquote {\bibinfo
  {title} {Discovery of the pure {P}olycyclic {A}romatic {H}ydrocarbon indene
  (c-{C}$_9${H}$_8$) with {GOTHAM} observations of {TMC}-1},}\ }\href {\doibase
  10.3847/2041-8213/abfd3a} {\bibfield  {journal} {\bibinfo  {journal}
  {Astrophys. J. Lett.}\ }\textbf {\bibinfo {volume} {913}},\ \bibinfo {eid}
  {L18} (\bibinfo {year} {2021})},\ \Eprint {http://arxiv.org/abs/2104.15117}
  {arXiv:2104.15117 [astro-ph.GA]} \BibitemShut {NoStop}%
\bibitem [{\citenamefont {Cernicharo}\ \emph {et~al.}(2021)\citenamefont
  {Cernicharo}, \citenamefont {Ag{\'u}ndez}, \citenamefont {Cabezas},
  \citenamefont {Tercero}, \citenamefont {Marcelino}, \citenamefont {Pardo},\
  and\ \citenamefont {de~Vicente}}]{Cernicharo2021}%
  \BibitemOpen
  \bibfield  {author} {\bibinfo {author} {\bibfnamefont {J.}~\bibnamefont
  {Cernicharo}}, \bibinfo {author} {\bibfnamefont {M.}~\bibnamefont
  {Ag{\'u}ndez}}, \bibinfo {author} {\bibfnamefont {C.}~\bibnamefont
  {Cabezas}}, \bibinfo {author} {\bibfnamefont {B.}~\bibnamefont {Tercero}},
  \bibinfo {author} {\bibfnamefont {N.}~\bibnamefont {Marcelino}}, \bibinfo
  {author} {\bibfnamefont {J.~R.}\ \bibnamefont {Pardo}}, \ and\ \bibinfo
  {author} {\bibfnamefont {P.}~\bibnamefont {de~Vicente}},\ }\bibfield  {title}
  {\enquote {\bibinfo {title} {Pure hydrocarbon cycles in tmc-1: Discovery of
  ethynyl cyclopropenylidene, cyclopentadiene, and indene},}\ }\href {\doibase
  10.1051/0004-6361/202141156} {\bibfield  {journal} {\bibinfo  {journal}
  {Astron. Astrophys.}\ }\textbf {\bibinfo {volume} {649}},\ \bibinfo {pages}
  {L15} (\bibinfo {year} {2021})}\BibitemShut {NoStop}%
\bibitem [{\citenamefont {Sita}\ \emph {et~al.}(2022)\citenamefont {Sita},
  \citenamefont {Changala}, \citenamefont {Xue}, \citenamefont {Burkhardt},
  \citenamefont {Shingledecker}, \citenamefont {Lee}, \citenamefont {Loomis},
  \citenamefont {Momjian}, \citenamefont {Siebert}, \citenamefont {Gupta},
  \citenamefont {Herbst}, \citenamefont {Remijan}, \citenamefont {McCarthy},
  \citenamefont {Cooke},\ and\ \citenamefont {McGuire}}]{Sita2022}%
  \BibitemOpen
  \bibfield  {author} {\bibinfo {author} {\bibfnamefont {M.~L.}\ \bibnamefont
  {Sita}}, \bibinfo {author} {\bibfnamefont {P.~B.}\ \bibnamefont {Changala}},
  \bibinfo {author} {\bibfnamefont {C.}~\bibnamefont {Xue}}, \bibinfo {author}
  {\bibfnamefont {A.~M.}\ \bibnamefont {Burkhardt}}, \bibinfo {author}
  {\bibfnamefont {C.~N.}\ \bibnamefont {Shingledecker}}, \bibinfo {author}
  {\bibfnamefont {K.~L.~K.}\ \bibnamefont {Lee}}, \bibinfo {author}
  {\bibfnamefont {R.~A.}\ \bibnamefont {Loomis}}, \bibinfo {author}
  {\bibfnamefont {E.}~\bibnamefont {Momjian}}, \bibinfo {author} {\bibfnamefont
  {M.~A.}\ \bibnamefont {Siebert}}, \bibinfo {author} {\bibfnamefont
  {D.}~\bibnamefont {Gupta}}, \bibinfo {author} {\bibfnamefont
  {E.}~\bibnamefont {Herbst}}, \bibinfo {author} {\bibfnamefont {A.~J.}\
  \bibnamefont {Remijan}}, \bibinfo {author} {\bibfnamefont {M.~C.}\
  \bibnamefont {McCarthy}}, \bibinfo {author} {\bibfnamefont {I.~R.}\
  \bibnamefont {Cooke}}, \ and\ \bibinfo {author} {\bibfnamefont {B.~A.}\
  \bibnamefont {McGuire}},\ }\bibfield  {title} {\enquote {\bibinfo {title}
  {Discovery of interstellar 2-cyanoindene (2-c9h7cn) in gotham observations of
  tmc-1},}\ }\href {\doibase 10.3847/2041-8213/ac92f4} {\bibfield  {journal}
  {\bibinfo  {journal} {Astrophys. J. Lett.}\ }\textbf {\bibinfo {volume}
  {938}},\ \bibinfo {pages} {L12} (\bibinfo {year} {2022})}\BibitemShut
  {NoStop}%
\bibitem [{\citenamefont {McGuire}\ \emph {et~al.}(2021)\citenamefont
  {McGuire}, \citenamefont {Loomis}, \citenamefont {Burkhardt}, \citenamefont
  {Lee}, \citenamefont {Shingledecker}, \citenamefont {Charnley}, \citenamefont
  {Cooke}, \citenamefont {Cordiner}, \citenamefont {Herbst}, \citenamefont
  {Kalenskii}, \citenamefont {Siebert}, \citenamefont {Willis}, \citenamefont
  {Xue}, \citenamefont {Remijan},\ and\ \citenamefont
  {McCarthy}}]{McGuire2021}%
  \BibitemOpen
  \bibfield  {author} {\bibinfo {author} {\bibfnamefont {B.~A.}\ \bibnamefont
  {McGuire}}, \bibinfo {author} {\bibfnamefont {R.~A.}\ \bibnamefont {Loomis}},
  \bibinfo {author} {\bibfnamefont {A.~M.}\ \bibnamefont {Burkhardt}}, \bibinfo
  {author} {\bibfnamefont {K.~L.~K.}\ \bibnamefont {Lee}}, \bibinfo {author}
  {\bibfnamefont {C.~N.}\ \bibnamefont {Shingledecker}}, \bibinfo {author}
  {\bibfnamefont {S.~B.}\ \bibnamefont {Charnley}}, \bibinfo {author}
  {\bibfnamefont {I.~R.}\ \bibnamefont {Cooke}}, \bibinfo {author}
  {\bibfnamefont {M.~A.}\ \bibnamefont {Cordiner}}, \bibinfo {author}
  {\bibfnamefont {E.}~\bibnamefont {Herbst}}, \bibinfo {author} {\bibfnamefont
  {S.}~\bibnamefont {Kalenskii}}, \bibinfo {author} {\bibfnamefont {M.~A.}\
  \bibnamefont {Siebert}}, \bibinfo {author} {\bibfnamefont {E.~R.}\
  \bibnamefont {Willis}}, \bibinfo {author} {\bibfnamefont {C.}~\bibnamefont
  {Xue}}, \bibinfo {author} {\bibfnamefont {A.~J.}\ \bibnamefont {Remijan}}, \
  and\ \bibinfo {author} {\bibfnamefont {M.~C.}\ \bibnamefont {McCarthy}},\
  }\bibfield  {title} {\enquote {\bibinfo {title} {Detection of two
  interstellar polycyclic aromatic hydrocarbons via spectral matched
  filtering},}\ }\href {\doibase 10.1126/science.abb7535} {\bibfield  {journal}
  {\bibinfo  {journal} {Science}\ }\textbf {\bibinfo {volume} {371}},\ \bibinfo
  {pages} {1265--1269} (\bibinfo {year} {2021})},\ \Eprint
  {http://arxiv.org/abs/2103.09984} {arXiv:2103.09984 [astro-ph.GA]}
  \BibitemShut {NoStop}%
\bibitem [{\citenamefont {Cernicharo}\ \emph {et~al.}(2024)\citenamefont
  {Cernicharo}, \citenamefont {Cabezas}, \citenamefont {Fuentetaja},
  \citenamefont {Agúndez}, \citenamefont {Tercero}, \citenamefont {Janeiro},
  \citenamefont {Juanes}, \citenamefont {Kaiser}, \citenamefont {Endo},
  \citenamefont {Steber}, \citenamefont {Pérez}, \citenamefont {Pérez},
  \citenamefont {Lesarri}, \citenamefont {Marcelino},\ and\ \citenamefont
  {de~Vicente}}]{Cernicharo2024}%
  \BibitemOpen
  \bibfield  {author} {\bibinfo {author} {\bibfnamefont {J.}~\bibnamefont
  {Cernicharo}}, \bibinfo {author} {\bibfnamefont {C.}~\bibnamefont {Cabezas}},
  \bibinfo {author} {\bibfnamefont {R.}~\bibnamefont {Fuentetaja}}, \bibinfo
  {author} {\bibfnamefont {M.}~\bibnamefont {Agúndez}}, \bibinfo {author}
  {\bibfnamefont {B.}~\bibnamefont {Tercero}}, \bibinfo {author} {\bibfnamefont
  {J.}~\bibnamefont {Janeiro}}, \bibinfo {author} {\bibfnamefont
  {M.}~\bibnamefont {Juanes}}, \bibinfo {author} {\bibfnamefont {R.~I.}\
  \bibnamefont {Kaiser}}, \bibinfo {author} {\bibfnamefont {Y.}~\bibnamefont
  {Endo}}, \bibinfo {author} {\bibfnamefont {A.~L.}\ \bibnamefont {Steber}},
  \bibinfo {author} {\bibfnamefont {D.}~\bibnamefont {Pérez}}, \bibinfo
  {author} {\bibfnamefont {C.}~\bibnamefont {Pérez}}, \bibinfo {author}
  {\bibfnamefont {A.}~\bibnamefont {Lesarri}}, \bibinfo {author} {\bibfnamefont
  {N.}~\bibnamefont {Marcelino}}, \ and\ \bibinfo {author} {\bibfnamefont
  {P.}~\bibnamefont {de~Vicente}},\ }\bibfield  {title} {\enquote {\bibinfo
  {title} {Discovery of two cyano derivatives of acenaphthylene (c$_{12}$h$_8$)
  in tmc-1 with the quijote line survey},}\ }\href {\doibase
  10.1051/0004-6361/202452196} {\bibfield  {journal} {\bibinfo  {journal}
  {Astron. Astrophys.}\ } (\bibinfo {year} {2024}),\
  10.1051/0004-6361/202452196},\ \Eprint {http://arxiv.org/abs/2409.19311}
  {arXiv:2409.19311 [astro-ph.GA]} \BibitemShut {NoStop}%
\bibitem [{\citenamefont {{Wenzel}}\ \emph {et~al.}(2024)\citenamefont
  {{Wenzel}}, \citenamefont {{Speak}}, \citenamefont {{Changala}},
  \citenamefont {{Willis}}, \citenamefont {{Burkhardt}}, \citenamefont
  {{Zhang}}, \citenamefont {{Bergin}}, \citenamefont {{Byrne}}, \citenamefont
  {{Charnley}}, \citenamefont {{Fried}}, \citenamefont {{Gupta}}, \citenamefont
  {{Herbst}}, \citenamefont {{Holdren}}, \citenamefont {{Lipnicky}},
  \citenamefont {{Loomis}}, \citenamefont {{Shingledecker}}, \citenamefont
  {{Xue}}, \citenamefont {{Remijan}}, \citenamefont {{Wendlandt}},
  \citenamefont {{McCarthy}}, \citenamefont {{Cooke}},\ and\ \citenamefont
  {{McGuire}}}]{Wenzel2024}%
  \BibitemOpen
  \bibfield  {author} {\bibinfo {author} {\bibfnamefont {G.}~\bibnamefont
  {{Wenzel}}}, \bibinfo {author} {\bibfnamefont {T.~H.}\ \bibnamefont
  {{Speak}}}, \bibinfo {author} {\bibfnamefont {P.~B.}\ \bibnamefont
  {{Changala}}}, \bibinfo {author} {\bibfnamefont {R.~H.~J.}\ \bibnamefont
  {{Willis}}}, \bibinfo {author} {\bibfnamefont {A.~M.}\ \bibnamefont
  {{Burkhardt}}}, \bibinfo {author} {\bibfnamefont {S.}~\bibnamefont
  {{Zhang}}}, \bibinfo {author} {\bibfnamefont {E.~A.}\ \bibnamefont
  {{Bergin}}}, \bibinfo {author} {\bibfnamefont {A.~N.}\ \bibnamefont
  {{Byrne}}}, \bibinfo {author} {\bibfnamefont {S.~B.}\ \bibnamefont
  {{Charnley}}}, \bibinfo {author} {\bibfnamefont {Z.~T.~P.}\ \bibnamefont
  {{Fried}}}, \bibinfo {author} {\bibfnamefont {H.}~\bibnamefont {{Gupta}}},
  \bibinfo {author} {\bibfnamefont {E.}~\bibnamefont {{Herbst}}}, \bibinfo
  {author} {\bibfnamefont {M.~S.}\ \bibnamefont {{Holdren}}}, \bibinfo {author}
  {\bibfnamefont {A.}~\bibnamefont {{Lipnicky}}}, \bibinfo {author}
  {\bibfnamefont {R.~A.}\ \bibnamefont {{Loomis}}}, \bibinfo {author}
  {\bibfnamefont {C.~N.}\ \bibnamefont {{Shingledecker}}}, \bibinfo {author}
  {\bibfnamefont {C.}~\bibnamefont {{Xue}}}, \bibinfo {author} {\bibfnamefont
  {A.~J.}\ \bibnamefont {{Remijan}}}, \bibinfo {author} {\bibfnamefont {A.~E.}\
  \bibnamefont {{Wendlandt}}}, \bibinfo {author} {\bibfnamefont {M.~C.}\
  \bibnamefont {{McCarthy}}}, \bibinfo {author} {\bibfnamefont {I.~R.}\
  \bibnamefont {{Cooke}}}, \ and\ \bibinfo {author} {\bibfnamefont {B.~A.}\
  \bibnamefont {{McGuire}}},\ }\bibfield  {title} {\enquote {\bibinfo {title}
  {{Detections of interstellar 2-cyanopyrene and 4-cyanopyrene in TMC-1}},}\
  }\href {\doibase 10.48550/arXiv.2410.00670} {\bibfield  {journal} {\bibinfo
  {journal} {arXiv e-prints}\ ,\ \bibinfo {eid} {arXiv:2410.00670}} (\bibinfo
  {year} {2024})},\ \Eprint {http://arxiv.org/abs/2410.00670} {arXiv:2410.00670
  [astro-ph.GA]} \BibitemShut {NoStop}%
\bibitem [{\citenamefont {Stockett}\ \emph {et~al.}(2023)\citenamefont
  {Stockett}, \citenamefont {Bull}, \citenamefont {Cederquist}, \citenamefont
  {Indrajith}, \citenamefont {Ji}, \citenamefont {Navarro~Navarrete},
  \citenamefont {Schmidt}, \citenamefont {Zettergren},\ and\ \citenamefont
  {Zhu}}]{Stockett2023}%
  \BibitemOpen
  \bibfield  {author} {\bibinfo {author} {\bibfnamefont {M.~H.}\ \bibnamefont
  {Stockett}}, \bibinfo {author} {\bibfnamefont {J.~N.}\ \bibnamefont {Bull}},
  \bibinfo {author} {\bibfnamefont {H.}~\bibnamefont {Cederquist}}, \bibinfo
  {author} {\bibfnamefont {S.}~\bibnamefont {Indrajith}}, \bibinfo {author}
  {\bibfnamefont {M.}~\bibnamefont {Ji}}, \bibinfo {author} {\bibfnamefont
  {J.~E.}\ \bibnamefont {Navarro~Navarrete}}, \bibinfo {author} {\bibfnamefont
  {H.~T.}\ \bibnamefont {Schmidt}}, \bibinfo {author} {\bibfnamefont
  {H.}~\bibnamefont {Zettergren}}, \ and\ \bibinfo {author} {\bibfnamefont
  {B.}~\bibnamefont {Zhu}},\ }\bibfield  {title} {\enquote {\bibinfo {title}
  {Efficient stabilization of cyanonaphthalene by fast radiative cooling and
  implications for the resilience of small pahs in interstellar clouds},}\
  }\href {\doibase 10.1038/s41467-023-36092-0} {\bibfield  {journal} {\bibinfo
  {journal} {Nat. Commun.}\ }\textbf {\bibinfo {volume} {14}},\ \bibinfo
  {pages} {395} (\bibinfo {year} {2023})}\BibitemShut {NoStop}%
\bibitem [{\citenamefont {L{\'e}ger}, \citenamefont {Boissel},\ and\
  \citenamefont {d'Hendecourt}(1988)}]{Leger1988}%
  \BibitemOpen
  \bibfield  {author} {\bibinfo {author} {\bibfnamefont {A.}~\bibnamefont
  {L{\'e}ger}}, \bibinfo {author} {\bibfnamefont {P.}~\bibnamefont {Boissel}},
  \ and\ \bibinfo {author} {\bibfnamefont {L.}~\bibnamefont {d'Hendecourt}},\
  }\bibfield  {title} {\enquote {\bibinfo {title} {Predicted fluorescence
  mechanism in highly isolated molecules: The poincar\'e fluorescence},}\
  }\href {\doibase 10.1103/PhysRevLett.60.921} {\bibfield  {journal} {\bibinfo
  {journal} {Phys. Rev. Lett.}\ }\textbf {\bibinfo {volume} {60}},\ \bibinfo
  {pages} {921--924} (\bibinfo {year} {1988})}\BibitemShut {NoStop}%
\bibitem [{\citenamefont {Boissel}\ \emph {et~al.}(1997)\citenamefont
  {Boissel}, \citenamefont {de~Parseval}, \citenamefont {Marty},\ and\
  \citenamefont {Lef{\`e}vre}}]{Boissel1997}%
  \BibitemOpen
  \bibfield  {author} {\bibinfo {author} {\bibfnamefont {P.}~\bibnamefont
  {Boissel}}, \bibinfo {author} {\bibfnamefont {P.}~\bibnamefont
  {de~Parseval}}, \bibinfo {author} {\bibfnamefont {P.}~\bibnamefont {Marty}},
  \ and\ \bibinfo {author} {\bibfnamefont {G.}~\bibnamefont {Lef{\`e}vre}},\
  }\bibfield  {title} {\enquote {\bibinfo {title} {Fragmentation of isolated
  ions by multiple photon absorption: A quantitative study},}\ }\href {\doibase
  10.1063/1.473545} {\bibfield  {journal} {\bibinfo  {journal} {J. Chem.
  Phys.}\ }\textbf {\bibinfo {volume} {106}},\ \bibinfo {pages} {4973--4984}
  (\bibinfo {year} {1997})},\ \Eprint
  {http://arxiv.org/abs/https://doi.org/10.1063/1.473545}
  {https://doi.org/10.1063/1.473545} \BibitemShut {NoStop}%
\bibitem [{\citenamefont {Martin}\ \emph {et~al.}(2013)\citenamefont {Martin},
  \citenamefont {Bernard}, \citenamefont {Br{\'e}dy}, \citenamefont {Concina},
  \citenamefont {Joblin}, \citenamefont {Ji}, \citenamefont {Ortega},\ and\
  \citenamefont {Chen}}]{Martin2013}%
  \BibitemOpen
  \bibfield  {author} {\bibinfo {author} {\bibfnamefont {S.}~\bibnamefont
  {Martin}}, \bibinfo {author} {\bibfnamefont {J.}~\bibnamefont {Bernard}},
  \bibinfo {author} {\bibfnamefont {R.}~\bibnamefont {Br{\'e}dy}}, \bibinfo
  {author} {\bibfnamefont {B.}~\bibnamefont {Concina}}, \bibinfo {author}
  {\bibfnamefont {C.}~\bibnamefont {Joblin}}, \bibinfo {author} {\bibfnamefont
  {M.}~\bibnamefont {Ji}}, \bibinfo {author} {\bibfnamefont {C.}~\bibnamefont
  {Ortega}}, \ and\ \bibinfo {author} {\bibfnamefont {L.}~\bibnamefont
  {Chen}},\ }\bibfield  {title} {\enquote {\bibinfo {title} {{Fast Radiative
  Cooling of Anthracene Observed in a Compact Electrostatic Storage Ring}},}\
  }\href {\doibase 10.1103/PhysRevLett.110.063003} {\bibfield  {journal}
  {\bibinfo  {journal} {Phys. Rev. Lett.}\ }\textbf {\bibinfo {volume} {110}},\
  \bibinfo {pages} {063003} (\bibinfo {year} {2013})}\BibitemShut {NoStop}%
\bibitem [{\citenamefont {Martin}\ \emph {et~al.}(2015)\citenamefont {Martin},
  \citenamefont {Ji}, \citenamefont {Bernard}, \citenamefont {Br\'edy},
  \citenamefont {Concina}, \citenamefont {Allouche}, \citenamefont {Joblin},
  \citenamefont {Ortega}, \citenamefont {Montagne}, \citenamefont {Cassimi},
  \citenamefont {Ngono-Ravache},\ and\ \citenamefont {Chen}}]{Martin2015}%
  \BibitemOpen
  \bibfield  {author} {\bibinfo {author} {\bibfnamefont {S.}~\bibnamefont
  {Martin}}, \bibinfo {author} {\bibfnamefont {M.}~\bibnamefont {Ji}}, \bibinfo
  {author} {\bibfnamefont {J.}~\bibnamefont {Bernard}}, \bibinfo {author}
  {\bibfnamefont {R.}~\bibnamefont {Br\'edy}}, \bibinfo {author} {\bibfnamefont
  {B.}~\bibnamefont {Concina}}, \bibinfo {author} {\bibfnamefont {A.~R.}\
  \bibnamefont {Allouche}}, \bibinfo {author} {\bibfnamefont {C.}~\bibnamefont
  {Joblin}}, \bibinfo {author} {\bibfnamefont {C.}~\bibnamefont {Ortega}},
  \bibinfo {author} {\bibfnamefont {G.}~\bibnamefont {Montagne}}, \bibinfo
  {author} {\bibfnamefont {A.}~\bibnamefont {Cassimi}}, \bibinfo {author}
  {\bibfnamefont {Y.}~\bibnamefont {Ngono-Ravache}}, \ and\ \bibinfo {author}
  {\bibfnamefont {L.}~\bibnamefont {Chen}},\ }\bibfield  {title} {\enquote
  {\bibinfo {title} {Fast radiative cooling of anthracene: Dependence on
  internal energy},}\ }\href {\doibase 10.1103/PhysRevA.92.053425} {\bibfield
  {journal} {\bibinfo  {journal} {Phys. Rev. A}\ }\textbf {\bibinfo {volume}
  {92}},\ \bibinfo {pages} {053425} (\bibinfo {year} {2015})}\BibitemShut
  {NoStop}%
\bibitem [{\citenamefont {Ji}\ \emph {et~al.}(2017)\citenamefont {Ji},
  \citenamefont {Bernard}, \citenamefont {Chen}, \citenamefont {Br{\'e}dy},
  \citenamefont {Ort{\'e}ga}, \citenamefont {Joblin}, \citenamefont {Cassimi},\
  and\ \citenamefont {Martin}}]{Ji2017}%
  \BibitemOpen
  \bibfield  {author} {\bibinfo {author} {\bibfnamefont {M.}~\bibnamefont
  {Ji}}, \bibinfo {author} {\bibfnamefont {J.}~\bibnamefont {Bernard}},
  \bibinfo {author} {\bibfnamefont {L.}~\bibnamefont {Chen}}, \bibinfo {author}
  {\bibfnamefont {R.}~\bibnamefont {Br{\'e}dy}}, \bibinfo {author}
  {\bibfnamefont {C.}~\bibnamefont {Ort{\'e}ga}}, \bibinfo {author}
  {\bibfnamefont {C.}~\bibnamefont {Joblin}}, \bibinfo {author} {\bibfnamefont
  {A.}~\bibnamefont {Cassimi}}, \ and\ \bibinfo {author} {\bibfnamefont
  {S.}~\bibnamefont {Martin}},\ }\bibfield  {title} {\enquote {\bibinfo {title}
  {Cooling of isolated anthracene cations probed with photons of different
  wavelengths in the mini-ring},}\ }\href {\doibase 10.1063/1.4973651}
  {\bibfield  {journal} {\bibinfo  {journal} {J. Chem. Phys.}\ }\textbf
  {\bibinfo {volume} {146}},\ \bibinfo {pages} {044301} (\bibinfo {year}
  {2017})},\ \Eprint {http://arxiv.org/abs/https://doi.org/10.1063/1.4973651}
  {https://doi.org/10.1063/1.4973651} \BibitemShut {NoStop}%
\bibitem [{\citenamefont {Saito}\ \emph {et~al.}(2020)\citenamefont {Saito},
  \citenamefont {Kubota}, \citenamefont {Yamasa}, \citenamefont {Suzuki},
  \citenamefont {Majima},\ and\ \citenamefont {Tsuchida}}]{Saito2020}%
  \BibitemOpen
  \bibfield  {author} {\bibinfo {author} {\bibfnamefont {M.}~\bibnamefont
  {Saito}}, \bibinfo {author} {\bibfnamefont {H.}~\bibnamefont {Kubota}},
  \bibinfo {author} {\bibfnamefont {K.}~\bibnamefont {Yamasa}}, \bibinfo
  {author} {\bibfnamefont {K.}~\bibnamefont {Suzuki}}, \bibinfo {author}
  {\bibfnamefont {T.}~\bibnamefont {Majima}}, \ and\ \bibinfo {author}
  {\bibfnamefont {H.}~\bibnamefont {Tsuchida}},\ }\bibfield  {title} {\enquote
  {\bibinfo {title} {Direct measurement of recurrent fluorescence emission from
  naphthalene ions},}\ }\href {\doibase 10.1103/PhysRevA.102.012820} {\bibfield
   {journal} {\bibinfo  {journal} {Phys. Rev. A}\ }\textbf {\bibinfo {volume}
  {102}},\ \bibinfo {pages} {012820} (\bibinfo {year} {2020})}\BibitemShut
  {NoStop}%
\bibitem [{\citenamefont {Stockett}\ \emph {et~al.}(2020)\citenamefont
  {Stockett}, \citenamefont {Bull}, \citenamefont {Buntine}, \citenamefont
  {Carrascosa}, \citenamefont {Ji}, \citenamefont {Kono}, \citenamefont
  {Schmidt},\ and\ \citenamefont {Zettergren}}]{Stockett2020b}%
  \BibitemOpen
  \bibfield  {author} {\bibinfo {author} {\bibfnamefont {M.~H.}\ \bibnamefont
  {Stockett}}, \bibinfo {author} {\bibfnamefont {J.~N.}\ \bibnamefont {Bull}},
  \bibinfo {author} {\bibfnamefont {J.~T.}\ \bibnamefont {Buntine}}, \bibinfo
  {author} {\bibfnamefont {E.}~\bibnamefont {Carrascosa}}, \bibinfo {author}
  {\bibfnamefont {M.}~\bibnamefont {Ji}}, \bibinfo {author} {\bibfnamefont
  {N.}~\bibnamefont {Kono}}, \bibinfo {author} {\bibfnamefont {H.~T.}\
  \bibnamefont {Schmidt}}, \ and\ \bibinfo {author} {\bibfnamefont
  {H.}~\bibnamefont {Zettergren}},\ }\bibfield  {title} {\enquote {\bibinfo
  {title} {Unimolecular fragmentation and radiative cooling of isolated {PAH}
  ions: A quantitative study},}\ }\href {\doibase 10.1063/5.0027773} {\bibfield
   {journal} {\bibinfo  {journal} {J. Chem. Phys.}\ }\textbf {\bibinfo {volume}
  {153}},\ \bibinfo {pages} {154303} (\bibinfo {year} {2020})},\ \Eprint
  {http://arxiv.org/abs/https://aip.scitation.org/doi/pdf/10.1063/5.0027773}
  {https://aip.scitation.org/doi/pdf/10.1063/5.0027773} \BibitemShut {NoStop}%
\bibitem [{\citenamefont {Lee}\ \emph {et~al.}(2023)\citenamefont {Lee},
  \citenamefont {Stockett}, \citenamefont {Ashworth}, \citenamefont
  {Navarro~Navarrete}, \citenamefont {Gougoula}, \citenamefont {Garg},
  \citenamefont {Ji}, \citenamefont {Zhu}, \citenamefont {Indrajith},
  \citenamefont {Zettergren}, \citenamefont {Schmidt},\ and\ \citenamefont
  {Bull}}]{Lee2023}%
  \BibitemOpen
  \bibfield  {author} {\bibinfo {author} {\bibfnamefont {J.~W.~L.}\
  \bibnamefont {Lee}}, \bibinfo {author} {\bibfnamefont {M.~H.}\ \bibnamefont
  {Stockett}}, \bibinfo {author} {\bibfnamefont {E.~K.}\ \bibnamefont
  {Ashworth}}, \bibinfo {author} {\bibfnamefont {J.~E.}\ \bibnamefont
  {Navarro~Navarrete}}, \bibinfo {author} {\bibfnamefont {E.}~\bibnamefont
  {Gougoula}}, \bibinfo {author} {\bibfnamefont {D.}~\bibnamefont {Garg}},
  \bibinfo {author} {\bibfnamefont {M.}~\bibnamefont {Ji}}, \bibinfo {author}
  {\bibfnamefont {B.}~\bibnamefont {Zhu}}, \bibinfo {author} {\bibfnamefont
  {S.}~\bibnamefont {Indrajith}}, \bibinfo {author} {\bibfnamefont
  {H.}~\bibnamefont {Zettergren}}, \bibinfo {author} {\bibfnamefont {H.~T.}\
  \bibnamefont {Schmidt}}, \ and\ \bibinfo {author} {\bibfnamefont {J.~N.}\
  \bibnamefont {Bull}},\ }\bibfield  {title} {\enquote {\bibinfo {title}
  {{Cooling dynamics of energized naphthalene and azulene radical cations}},}\
  }\href {\doibase 10.1063/5.0147456} {\bibfield  {journal} {\bibinfo
  {journal} {J. Chem. Phys.}\ }\textbf {\bibinfo {volume} {158}},\ \bibinfo
  {pages} {174305} (\bibinfo {year} {2023})},\ \bibinfo {note} {174305},\
  \Eprint
  {http://arxiv.org/abs/https://pubs.aip.org/aip/jcp/article-pdf/doi/10.1063/5.0147456/17145753/174305\_1\_5.0147456.pdf}
  {https://pubs.aip.org/aip/jcp/article-pdf/doi/10.1063/5.0147456/17145753/174305\_1\_5.0147456.pdf}
  \BibitemShut {NoStop}%
\bibitem [{\citenamefont {Navarro~Navarrete}\ \emph {et~al.}(2023)\citenamefont
  {Navarro~Navarrete}, \citenamefont {Bull}, \citenamefont {Cederquist},
  \citenamefont {Indrajith}, \citenamefont {Ji}, \citenamefont {Schmidt},
  \citenamefont {Zettergren}, \citenamefont {Zhu},\ and\ \citenamefont
  {Stockett}}]{NavarroNavarrete2023}%
  \BibitemOpen
  \bibfield  {author} {\bibinfo {author} {\bibfnamefont {J.~E.}\ \bibnamefont
  {Navarro~Navarrete}}, \bibinfo {author} {\bibfnamefont {J.~N.}\ \bibnamefont
  {Bull}}, \bibinfo {author} {\bibfnamefont {H.}~\bibnamefont {Cederquist}},
  \bibinfo {author} {\bibfnamefont {S.}~\bibnamefont {Indrajith}}, \bibinfo
  {author} {\bibfnamefont {M.}~\bibnamefont {Ji}}, \bibinfo {author}
  {\bibfnamefont {H.}~\bibnamefont {Schmidt}}, \bibinfo {author} {\bibfnamefont
  {H.}~\bibnamefont {Zettergren}}, \bibinfo {author} {\bibfnamefont
  {B.}~\bibnamefont {Zhu}}, \ and\ \bibinfo {author} {\bibfnamefont {M.~H.}\
  \bibnamefont {Stockett}},\ }\bibfield  {title} {\enquote {\bibinfo {title}
  {Experimental radiative cooling rates of a polycyclic aromatic hydrocarbon
  cation},}\ }\href {\doibase 10.1039/D3FD00005B} {\bibfield  {journal}
  {\bibinfo  {journal} {Faraday Discuss.}\ }\textbf {\bibinfo {volume} {245}},\
  \bibinfo {pages} {352--367} (\bibinfo {year} {2023})}\BibitemShut {NoStop}%
\bibitem [{\citenamefont {Bernard}\ \emph {et~al.}(2023)\citenamefont
  {Bernard}, \citenamefont {Ji}, \citenamefont {Indrajith}, \citenamefont
  {Stockett}, \citenamefont {Navarro~Navarrete}, \citenamefont {Kono},
  \citenamefont {Cederquist}, \citenamefont {Martin}, \citenamefont {Schmidt},\
  and\ \citenamefont {Zettergren}}]{Bernard2023}%
  \BibitemOpen
  \bibfield  {author} {\bibinfo {author} {\bibfnamefont {J.}~\bibnamefont
  {Bernard}}, \bibinfo {author} {\bibfnamefont {M.}~\bibnamefont {Ji}},
  \bibinfo {author} {\bibfnamefont {S.}~\bibnamefont {Indrajith}}, \bibinfo
  {author} {\bibfnamefont {M.~H.}\ \bibnamefont {Stockett}}, \bibinfo {author}
  {\bibfnamefont {J.~E.}\ \bibnamefont {Navarro~Navarrete}}, \bibinfo {author}
  {\bibfnamefont {N.}~\bibnamefont {Kono}}, \bibinfo {author} {\bibfnamefont
  {H.}~\bibnamefont {Cederquist}}, \bibinfo {author} {\bibfnamefont
  {S.}~\bibnamefont {Martin}}, \bibinfo {author} {\bibfnamefont
  {H.}~\bibnamefont {Schmidt}}, \ and\ \bibinfo {author} {\bibfnamefont
  {H.}~\bibnamefont {Zettergren}},\ }\bibfield  {title} {\enquote {\bibinfo
  {title} {Efficient radiative cooling of tetracene cations
  {C}$_{18}${H}$_{12}^+$ : absolute recurrent fluorescence rates as a function
  of internal energy},}\ }\href {\doibase 10.1039/D3CP00424D} {\bibfield
  {journal} {\bibinfo  {journal} {Phys. Chem. Chem. Phys.}\ }\textbf {\bibinfo
  {volume} {25}},\ \bibinfo {pages} {10726--10740} (\bibinfo {year}
  {2023})}\BibitemShut {NoStop}%
\bibitem [{\citenamefont {Zhu}\ \emph {et~al.}(2022)\citenamefont {Zhu},
  \citenamefont {Bull}, \citenamefont {Ji}, \citenamefont {Zettergren},\ and\
  \citenamefont {Stockett}}]{Zhu2022}%
  \BibitemOpen
  \bibfield  {author} {\bibinfo {author} {\bibfnamefont {B.}~\bibnamefont
  {Zhu}}, \bibinfo {author} {\bibfnamefont {J.~N.}\ \bibnamefont {Bull}},
  \bibinfo {author} {\bibfnamefont {M.}~\bibnamefont {Ji}}, \bibinfo {author}
  {\bibfnamefont {H.}~\bibnamefont {Zettergren}}, \ and\ \bibinfo {author}
  {\bibfnamefont {M.~H.}\ \bibnamefont {Stockett}},\ }\bibfield  {title}
  {\enquote {\bibinfo {title} {Radiative cooling rates of substituted {PAH}
  ions},}\ }\href {\doibase 10.1063/5.0089687} {\bibfield  {journal} {\bibinfo
  {journal} {J. Chem. Phys.}\ }\textbf {\bibinfo {volume} {157}},\ \bibinfo
  {pages} {044303} (\bibinfo {year} {2022})},\ \Eprint
  {http://arxiv.org/abs/https://doi.org/10.1063/5.0089687}
  {https://doi.org/10.1063/5.0089687} \BibitemShut {NoStop}%
\bibitem [{\citenamefont {Thomas}\ \emph {et~al.}(2011)\citenamefont {Thomas},
  \citenamefont {Schmidt}, \citenamefont {Andler}, \citenamefont
  {Bj{\"o}rkhage}, \citenamefont {Blom}, \citenamefont {Br{\"a}nnholm},
  \citenamefont {B{\"a}ckstr{\"o}m}, \citenamefont {Danared}, \citenamefont
  {Das}, \citenamefont {Haag}, \citenamefont {Halld{\'e}n}, \citenamefont
  {Hellberg}, \citenamefont {Holm}, \citenamefont {Johansson}, \citenamefont
  {K{\"a}llberg}, \citenamefont {K{\"a}llersj{\"o}}, \citenamefont {Larsson},
  \citenamefont {Leontein}, \citenamefont {Liljeby}, \citenamefont
  {L{\"o}fgren}, \citenamefont {Malm}, \citenamefont {Mannervik}, \citenamefont
  {Masuda}, \citenamefont {Misra}, \citenamefont {Orb{\'a}n}, \citenamefont
  {Pa{\'a}l}, \citenamefont {Reinhed}, \citenamefont {Rensfelt}, \citenamefont
  {Ros{\'e}n}, \citenamefont {Schmidt}, \citenamefont {Seitz}, \citenamefont
  {Simonsson}, \citenamefont {Weimer}, \citenamefont {Zettergren},\ and\
  \citenamefont {Cederquist}}]{Thomas2011}%
  \BibitemOpen
  \bibfield  {author} {\bibinfo {author} {\bibfnamefont {R.~D.}\ \bibnamefont
  {Thomas}}, \bibinfo {author} {\bibfnamefont {H.~T.}\ \bibnamefont {Schmidt}},
  \bibinfo {author} {\bibfnamefont {G.}~\bibnamefont {Andler}}, \bibinfo
  {author} {\bibfnamefont {M.}~\bibnamefont {Bj{\"o}rkhage}}, \bibinfo {author}
  {\bibfnamefont {M.}~\bibnamefont {Blom}}, \bibinfo {author} {\bibfnamefont
  {L.}~\bibnamefont {Br{\"a}nnholm}}, \bibinfo {author} {\bibfnamefont
  {E.}~\bibnamefont {B{\"a}ckstr{\"o}m}}, \bibinfo {author} {\bibfnamefont
  {H.}~\bibnamefont {Danared}}, \bibinfo {author} {\bibfnamefont
  {S.}~\bibnamefont {Das}}, \bibinfo {author} {\bibfnamefont {N.}~\bibnamefont
  {Haag}}, \bibinfo {author} {\bibfnamefont {P.}~\bibnamefont {Halld{\'e}n}},
  \bibinfo {author} {\bibfnamefont {F.}~\bibnamefont {Hellberg}}, \bibinfo
  {author} {\bibfnamefont {A.~I.~S.}\ \bibnamefont {Holm}}, \bibinfo {author}
  {\bibfnamefont {H.~A.~B.}\ \bibnamefont {Johansson}}, \bibinfo {author}
  {\bibfnamefont {A.}~\bibnamefont {K{\"a}llberg}}, \bibinfo {author}
  {\bibfnamefont {G.}~\bibnamefont {K{\"a}llersj{\"o}}}, \bibinfo {author}
  {\bibfnamefont {M.}~\bibnamefont {Larsson}}, \bibinfo {author} {\bibfnamefont
  {S.}~\bibnamefont {Leontein}}, \bibinfo {author} {\bibfnamefont
  {L.}~\bibnamefont {Liljeby}}, \bibinfo {author} {\bibfnamefont
  {P.}~\bibnamefont {L{\"o}fgren}}, \bibinfo {author} {\bibfnamefont
  {B.}~\bibnamefont {Malm}}, \bibinfo {author} {\bibfnamefont {S.}~\bibnamefont
  {Mannervik}}, \bibinfo {author} {\bibfnamefont {M.}~\bibnamefont {Masuda}},
  \bibinfo {author} {\bibfnamefont {D.}~\bibnamefont {Misra}}, \bibinfo
  {author} {\bibfnamefont {A.}~\bibnamefont {Orb{\'a}n}}, \bibinfo {author}
  {\bibfnamefont {A.}~\bibnamefont {Pa{\'a}l}}, \bibinfo {author}
  {\bibfnamefont {P.}~\bibnamefont {Reinhed}}, \bibinfo {author} {\bibfnamefont
  {K.-G.}\ \bibnamefont {Rensfelt}}, \bibinfo {author} {\bibfnamefont
  {S.}~\bibnamefont {Ros{\'e}n}}, \bibinfo {author} {\bibfnamefont
  {K.}~\bibnamefont {Schmidt}}, \bibinfo {author} {\bibfnamefont
  {F.}~\bibnamefont {Seitz}}, \bibinfo {author} {\bibfnamefont
  {A.}~\bibnamefont {Simonsson}}, \bibinfo {author} {\bibfnamefont
  {J.}~\bibnamefont {Weimer}}, \bibinfo {author} {\bibfnamefont
  {H.}~\bibnamefont {Zettergren}}, \ and\ \bibinfo {author} {\bibfnamefont
  {H.}~\bibnamefont {Cederquist}},\ }\bibfield  {title} {\enquote {\bibinfo
  {title} {{The {D}ouble {E}lectro{S}tatic {I}on {R}ing {E}xp{E}riment: A
  unique cryogenic electrostatic storage ring for merged ion-beams studies}},}\
  }\href {\doibase 10.1063/1.3602928} {\bibfield  {journal} {\bibinfo
  {journal} {Rev. Sci. Instrum.}\ }\textbf {\bibinfo {volume} {82}},\ \bibinfo
  {eid} {065112} (\bibinfo {year} {2011})}\BibitemShut {NoStop}%
\bibitem [{\citenamefont {Schmidt}\ \emph {et~al.}(2013)\citenamefont
  {Schmidt}, \citenamefont {Thomas}, \citenamefont {Gatchell}, \citenamefont
  {Ros{\'e}n}, \citenamefont {Reinhed}, \citenamefont {L{\"o}fgren},
  \citenamefont {Br{\"a}nnholm}, \citenamefont {Blom}, \citenamefont
  {Bj{\"o}rkhage}, \citenamefont {B{\"a}ckstr{\"o}m}, \citenamefont
  {Alexander}, \citenamefont {Leontein}, \citenamefont {Hanstorp},
  \citenamefont {Zettergren}, \citenamefont {Liljeby}, \citenamefont
  {K{\"a}llberg}, \citenamefont {Simonsson}, \citenamefont {Hellberg},
  \citenamefont {Mannervik}, \citenamefont {Larsson}, \citenamefont {Geppert},
  \citenamefont {Rensfelt}, \citenamefont {Danared}, \citenamefont {Pa{\'a}l},
  \citenamefont {Masuda}, \citenamefont {Halld{\'e}n}, \citenamefont {Andler},
  \citenamefont {Stockett}, \citenamefont {Chen}, \citenamefont
  {K{\"a}llersj{\"o}}, \citenamefont {Weimer}, \citenamefont {Hansen},
  \citenamefont {Hartman},\ and\ \citenamefont {Cederquist}}]{Schmidt2013}%
  \BibitemOpen
  \bibfield  {author} {\bibinfo {author} {\bibfnamefont {H.~T.}\ \bibnamefont
  {Schmidt}}, \bibinfo {author} {\bibfnamefont {R.~D.}\ \bibnamefont {Thomas}},
  \bibinfo {author} {\bibfnamefont {M.}~\bibnamefont {Gatchell}}, \bibinfo
  {author} {\bibfnamefont {S.}~\bibnamefont {Ros{\'e}n}}, \bibinfo {author}
  {\bibfnamefont {P.}~\bibnamefont {Reinhed}}, \bibinfo {author} {\bibfnamefont
  {P.}~\bibnamefont {L{\"o}fgren}}, \bibinfo {author} {\bibfnamefont
  {L.}~\bibnamefont {Br{\"a}nnholm}}, \bibinfo {author} {\bibfnamefont
  {M.}~\bibnamefont {Blom}}, \bibinfo {author} {\bibfnamefont {M.}~\bibnamefont
  {Bj{\"o}rkhage}}, \bibinfo {author} {\bibfnamefont {E.}~\bibnamefont
  {B{\"a}ckstr{\"o}m}}, \bibinfo {author} {\bibfnamefont {J.~D.}\ \bibnamefont
  {Alexander}}, \bibinfo {author} {\bibfnamefont {S.}~\bibnamefont {Leontein}},
  \bibinfo {author} {\bibfnamefont {D.}~\bibnamefont {Hanstorp}}, \bibinfo
  {author} {\bibfnamefont {H.}~\bibnamefont {Zettergren}}, \bibinfo {author}
  {\bibfnamefont {L.}~\bibnamefont {Liljeby}}, \bibinfo {author} {\bibfnamefont
  {A.}~\bibnamefont {K{\"a}llberg}}, \bibinfo {author} {\bibfnamefont
  {A.}~\bibnamefont {Simonsson}}, \bibinfo {author} {\bibfnamefont
  {F.}~\bibnamefont {Hellberg}}, \bibinfo {author} {\bibfnamefont
  {S.}~\bibnamefont {Mannervik}}, \bibinfo {author} {\bibfnamefont
  {M.}~\bibnamefont {Larsson}}, \bibinfo {author} {\bibfnamefont {W.~D.}\
  \bibnamefont {Geppert}}, \bibinfo {author} {\bibfnamefont {K.~G.}\
  \bibnamefont {Rensfelt}}, \bibinfo {author} {\bibfnamefont {H.}~\bibnamefont
  {Danared}}, \bibinfo {author} {\bibfnamefont {A.}~\bibnamefont {Pa{\'a}l}},
  \bibinfo {author} {\bibfnamefont {M.}~\bibnamefont {Masuda}}, \bibinfo
  {author} {\bibfnamefont {P.}~\bibnamefont {Halld{\'e}n}}, \bibinfo {author}
  {\bibfnamefont {G.}~\bibnamefont {Andler}}, \bibinfo {author} {\bibfnamefont
  {M.~H.}\ \bibnamefont {Stockett}}, \bibinfo {author} {\bibfnamefont
  {T.}~\bibnamefont {Chen}}, \bibinfo {author} {\bibfnamefont {G.}~\bibnamefont
  {K{\"a}llersj{\"o}}}, \bibinfo {author} {\bibfnamefont {J.}~\bibnamefont
  {Weimer}}, \bibinfo {author} {\bibfnamefont {K.}~\bibnamefont {Hansen}},
  \bibinfo {author} {\bibfnamefont {H.}~\bibnamefont {Hartman}}, \ and\
  \bibinfo {author} {\bibfnamefont {H.}~\bibnamefont {Cederquist}},\ }\bibfield
   {title} {\enquote {\bibinfo {title} {{First storage of ion beams in the
  Double Electrostatic Ion-Ring Experiment: DESIREE}},}\ }\href {\doibase
  10.1063/1.4807702} {\bibfield  {journal} {\bibinfo  {journal} {Rev. Sci.
  Instrum.}\ }\textbf {\bibinfo {volume} {84}},\ \bibinfo {eid} {055115}
  (\bibinfo {year} {2013})}\BibitemShut {NoStop}%
\bibitem [{\citenamefont {Gibson}\ \emph {et~al.}(2021)\citenamefont {Gibson},
  \citenamefont {Hickstein}, \citenamefont {Yurchak}, \citenamefont {Ryazanov},
  \citenamefont {Das},\ and\ \citenamefont {Shih}}]{pyabel}%
  \BibitemOpen
  \bibfield  {author} {\bibinfo {author} {\bibfnamefont {S.}~\bibnamefont
  {Gibson}}, \bibinfo {author} {\bibfnamefont {D.~D.}\ \bibnamefont
  {Hickstein}}, \bibinfo {author} {\bibfnamefont {R.}~\bibnamefont {Yurchak}},
  \bibinfo {author} {\bibfnamefont {M.}~\bibnamefont {Ryazanov}}, \bibinfo
  {author} {\bibfnamefont {D.}~\bibnamefont {Das}}, \ and\ \bibinfo {author}
  {\bibfnamefont {G.}~\bibnamefont {Shih}},\ }\href {\doibase
  10.5281/zenodo.4690660} {\enquote {\bibinfo {title} {Pyabel: v0.8.4
  https://doi.org/10.5281/zenodo.7438595},}\ } (\bibinfo {year}
  {2021})\BibitemShut {NoStop}%
\bibitem [{\citenamefont {Frisch}\ \emph {et~al.}(2016)\citenamefont {Frisch},
  \citenamefont {Trucks}, \citenamefont {Schlegel}, \citenamefont {Scuseria},
  \citenamefont {Robb}, \citenamefont {Cheeseman}, \citenamefont {Scalmani},
  \citenamefont {Barone}, \citenamefont {Mennucci}, \citenamefont {Petersson},
  \citenamefont {Nakatsuji}, \citenamefont {Caricato}, \citenamefont {Li},
  \citenamefont {Hratchian}, \citenamefont {Izmaylov}, \citenamefont {Bloino},
  \citenamefont {Zheng}, \citenamefont {Sonnenberg}, \citenamefont {Hada},
  \citenamefont {Ehara}, \citenamefont {Toyota}, \citenamefont {Fukuda},
  \citenamefont {Hasegawa}, \citenamefont {Ishida}, \citenamefont {Nakajima},
  \citenamefont {Honda}, \citenamefont {Kitao}, \citenamefont {Nakai},
  \citenamefont {Vreven}, \citenamefont {Montgomery}, \citenamefont {Peralta},
  \citenamefont {Ogliaro}, \citenamefont {Bearpark}, \citenamefont {Heyd},
  \citenamefont {Brothers}, \citenamefont {Kudin}, \citenamefont {Staroverov},
  \citenamefont {Kobayashi}, \citenamefont {Normand}, \citenamefont
  {Raghavachari}, \citenamefont {Rendell}, \citenamefont {Burant},
  \citenamefont {Iyengar}, \citenamefont {Tomasi}, \citenamefont {Cossi},
  \citenamefont {Rega}, \citenamefont {Millam}, \citenamefont {Klene},
  \citenamefont {Knox}, \citenamefont {Cross}, \citenamefont {Bakken},
  \citenamefont {Adamo}, \citenamefont {Jaramillo}, \citenamefont {Gomperts},
  \citenamefont {Stratmann}, \citenamefont {Yazyev}, \citenamefont {Austin},
  \citenamefont {Cammi}, \citenamefont {Pomelli}, \citenamefont {Ochterski},
  \citenamefont {Martin}, \citenamefont {Morokuma}, \citenamefont {Zakrzewski},
  \citenamefont {Voth}, \citenamefont {Salvador}, \citenamefont {Dannenberg},
  \citenamefont {Dapprich}, \citenamefont {Daniels}, \citenamefont {Farkas},
  \citenamefont {Foresman}, \citenamefont {Ortiz}, \citenamefont {Cioslowski},\
  and\ \citenamefont {Fox}}]{g16}%
  \BibitemOpen
  \bibfield  {author} {\bibinfo {author} {\bibfnamefont {M.~J.}\ \bibnamefont
  {Frisch}}, \bibinfo {author} {\bibfnamefont {G.~W.}\ \bibnamefont {Trucks}},
  \bibinfo {author} {\bibfnamefont {H.~B.}\ \bibnamefont {Schlegel}}, \bibinfo
  {author} {\bibfnamefont {G.~E.}\ \bibnamefont {Scuseria}}, \bibinfo {author}
  {\bibfnamefont {M.~A.}\ \bibnamefont {Robb}}, \bibinfo {author}
  {\bibfnamefont {J.~R.}\ \bibnamefont {Cheeseman}}, \bibinfo {author}
  {\bibfnamefont {G.}~\bibnamefont {Scalmani}}, \bibinfo {author}
  {\bibfnamefont {V.}~\bibnamefont {Barone}}, \bibinfo {author} {\bibfnamefont
  {B.}~\bibnamefont {Mennucci}}, \bibinfo {author} {\bibfnamefont {G.~A.}\
  \bibnamefont {Petersson}}, \bibinfo {author} {\bibfnamefont {H.}~\bibnamefont
  {Nakatsuji}}, \bibinfo {author} {\bibfnamefont {M.}~\bibnamefont {Caricato}},
  \bibinfo {author} {\bibfnamefont {X.}~\bibnamefont {Li}}, \bibinfo {author}
  {\bibfnamefont {H.~P.}\ \bibnamefont {Hratchian}}, \bibinfo {author}
  {\bibfnamefont {A.~F.}\ \bibnamefont {Izmaylov}}, \bibinfo {author}
  {\bibfnamefont {J.}~\bibnamefont {Bloino}}, \bibinfo {author} {\bibfnamefont
  {G.}~\bibnamefont {Zheng}}, \bibinfo {author} {\bibfnamefont {J.~L.}\
  \bibnamefont {Sonnenberg}}, \bibinfo {author} {\bibfnamefont
  {M.}~\bibnamefont {Hada}}, \bibinfo {author} {\bibfnamefont {M.}~\bibnamefont
  {Ehara}}, \bibinfo {author} {\bibfnamefont {K.}~\bibnamefont {Toyota}},
  \bibinfo {author} {\bibfnamefont {R.}~\bibnamefont {Fukuda}}, \bibinfo
  {author} {\bibfnamefont {J.}~\bibnamefont {Hasegawa}}, \bibinfo {author}
  {\bibfnamefont {M.}~\bibnamefont {Ishida}}, \bibinfo {author} {\bibfnamefont
  {T.}~\bibnamefont {Nakajima}}, \bibinfo {author} {\bibfnamefont
  {Y.}~\bibnamefont {Honda}}, \bibinfo {author} {\bibfnamefont
  {O.}~\bibnamefont {Kitao}}, \bibinfo {author} {\bibfnamefont
  {H.}~\bibnamefont {Nakai}}, \bibinfo {author} {\bibfnamefont
  {T.}~\bibnamefont {Vreven}}, \bibinfo {author} {\bibfnamefont {J.~A.}\
  \bibnamefont {Montgomery}, \bibfnamefont {{Jr.}}}, \bibinfo {author}
  {\bibfnamefont {J.~E.}\ \bibnamefont {Peralta}}, \bibinfo {author}
  {\bibfnamefont {F.}~\bibnamefont {Ogliaro}}, \bibinfo {author} {\bibfnamefont
  {M.}~\bibnamefont {Bearpark}}, \bibinfo {author} {\bibfnamefont {J.~J.}\
  \bibnamefont {Heyd}}, \bibinfo {author} {\bibfnamefont {E.}~\bibnamefont
  {Brothers}}, \bibinfo {author} {\bibfnamefont {K.~N.}\ \bibnamefont {Kudin}},
  \bibinfo {author} {\bibfnamefont {V.~N.}\ \bibnamefont {Staroverov}},
  \bibinfo {author} {\bibfnamefont {R.}~\bibnamefont {Kobayashi}}, \bibinfo
  {author} {\bibfnamefont {J.}~\bibnamefont {Normand}}, \bibinfo {author}
  {\bibfnamefont {K.}~\bibnamefont {Raghavachari}}, \bibinfo {author}
  {\bibfnamefont {A.}~\bibnamefont {Rendell}}, \bibinfo {author} {\bibfnamefont
  {J.~C.}\ \bibnamefont {Burant}}, \bibinfo {author} {\bibfnamefont {S.~S.}\
  \bibnamefont {Iyengar}}, \bibinfo {author} {\bibfnamefont {J.}~\bibnamefont
  {Tomasi}}, \bibinfo {author} {\bibfnamefont {M.}~\bibnamefont {Cossi}},
  \bibinfo {author} {\bibfnamefont {N.}~\bibnamefont {Rega}}, \bibinfo {author}
  {\bibfnamefont {J.~M.}\ \bibnamefont {Millam}}, \bibinfo {author}
  {\bibfnamefont {M.}~\bibnamefont {Klene}}, \bibinfo {author} {\bibfnamefont
  {J.~E.}\ \bibnamefont {Knox}}, \bibinfo {author} {\bibfnamefont {J.~B.}\
  \bibnamefont {Cross}}, \bibinfo {author} {\bibfnamefont {V.}~\bibnamefont
  {Bakken}}, \bibinfo {author} {\bibfnamefont {C.}~\bibnamefont {Adamo}},
  \bibinfo {author} {\bibfnamefont {J.}~\bibnamefont {Jaramillo}}, \bibinfo
  {author} {\bibfnamefont {R.}~\bibnamefont {Gomperts}}, \bibinfo {author}
  {\bibfnamefont {R.~E.}\ \bibnamefont {Stratmann}}, \bibinfo {author}
  {\bibfnamefont {O.}~\bibnamefont {Yazyev}}, \bibinfo {author} {\bibfnamefont
  {A.~J.}\ \bibnamefont {Austin}}, \bibinfo {author} {\bibfnamefont
  {R.}~\bibnamefont {Cammi}}, \bibinfo {author} {\bibfnamefont
  {C.}~\bibnamefont {Pomelli}}, \bibinfo {author} {\bibfnamefont {J.~W.}\
  \bibnamefont {Ochterski}}, \bibinfo {author} {\bibfnamefont {R.~L.}\
  \bibnamefont {Martin}}, \bibinfo {author} {\bibfnamefont {K.}~\bibnamefont
  {Morokuma}}, \bibinfo {author} {\bibfnamefont {V.~G.}\ \bibnamefont
  {Zakrzewski}}, \bibinfo {author} {\bibfnamefont {G.~A.}\ \bibnamefont
  {Voth}}, \bibinfo {author} {\bibfnamefont {P.}~\bibnamefont {Salvador}},
  \bibinfo {author} {\bibfnamefont {J.~J.}\ \bibnamefont {Dannenberg}},
  \bibinfo {author} {\bibfnamefont {S.}~\bibnamefont {Dapprich}}, \bibinfo
  {author} {\bibfnamefont {A.~D.}\ \bibnamefont {Daniels}}, \bibinfo {author}
  {\bibfnamefont {{\"O}.}~\bibnamefont {Farkas}}, \bibinfo {author}
  {\bibfnamefont {J.~B.}\ \bibnamefont {Foresman}}, \bibinfo {author}
  {\bibfnamefont {J.~V.}\ \bibnamefont {Ortiz}}, \bibinfo {author}
  {\bibfnamefont {J.}~\bibnamefont {Cioslowski}}, \ and\ \bibinfo {author}
  {\bibfnamefont {D.~J.}\ \bibnamefont {Fox}},\ }\href@noop {} {\enquote
  {\bibinfo {title} {Gaussian~16 {R}evision {B}.01},}\ } (\bibinfo {year}
  {2016}),\ \bibinfo {note} {gaussian Inc. Wallingford CT 2016}\BibitemShut
  {NoStop}%
\bibitem [{\citenamefont {Chandrasekaran}\ \emph {et~al.}(2014)\citenamefont
  {Chandrasekaran}, \citenamefont {Kafle}, \citenamefont {Prabhakaran},
  \citenamefont {Heber}, \citenamefont {Rappaport}, \citenamefont {Rubinstein},
  \citenamefont {Schwalm}, \citenamefont {Toker},\ and\ \citenamefont
  {Zajfman}}]{Chandrasekaran2014}%
  \BibitemOpen
  \bibfield  {author} {\bibinfo {author} {\bibfnamefont {V.}~\bibnamefont
  {Chandrasekaran}}, \bibinfo {author} {\bibfnamefont {B.}~\bibnamefont
  {Kafle}}, \bibinfo {author} {\bibfnamefont {A.}~\bibnamefont {Prabhakaran}},
  \bibinfo {author} {\bibfnamefont {O.}~\bibnamefont {Heber}}, \bibinfo
  {author} {\bibfnamefont {M.}~\bibnamefont {Rappaport}}, \bibinfo {author}
  {\bibfnamefont {H.}~\bibnamefont {Rubinstein}}, \bibinfo {author}
  {\bibfnamefont {D.}~\bibnamefont {Schwalm}}, \bibinfo {author} {\bibfnamefont
  {Y.}~\bibnamefont {Toker}}, \ and\ \bibinfo {author} {\bibfnamefont
  {D.}~\bibnamefont {Zajfman}},\ }\bibfield  {title} {\enquote {\bibinfo
  {title} {Determination of absolute recurrent fluorescence rate coefficients
  for {C}$_6^-$},}\ }\href@noop {} {\bibfield  {journal} {\bibinfo  {journal}
  {J. Phys. Chem. Lett.}\ }\textbf {\bibinfo {volume} {5}},\ \bibinfo {pages}
  {4078--4082} (\bibinfo {year} {2014})}\BibitemShut {NoStop}%
\bibitem [{\citenamefont {Beyer}\ and\ \citenamefont
  {Swinehart}(1973)}]{Beyer1973}%
  \BibitemOpen
  \bibfield  {author} {\bibinfo {author} {\bibfnamefont {T.}~\bibnamefont
  {Beyer}}\ and\ \bibinfo {author} {\bibfnamefont {D.~F.}\ \bibnamefont
  {Swinehart}},\ }\bibfield  {title} {\enquote {\bibinfo {title} {Algorithm
  448: Number of multiply-restricted partitions},}\ }\href {\doibase
  10.1145/362248.362275} {\bibfield  {journal} {\bibinfo  {journal} {Commun.
  ACM}\ }\textbf {\bibinfo {volume} {16}},\ \bibinfo {pages} {379} (\bibinfo
  {year} {1973})}\BibitemShut {NoStop}%
\bibitem [{\citenamefont {Bull}\ \emph {et~al.}(2019)\citenamefont {Bull},
  \citenamefont {Scholz}, \citenamefont {Carrascosa}, \citenamefont
  {Kristiansson}, \citenamefont {Eklund}, \citenamefont {Punnakayathil},
  \citenamefont {de~Ruette}, \citenamefont {Zettergren}, \citenamefont
  {Schmidt}, \citenamefont {Cederquist},\ and\ \citenamefont
  {Stockett}}]{Bull2019a}%
  \BibitemOpen
  \bibfield  {author} {\bibinfo {author} {\bibfnamefont {J.~N.}\ \bibnamefont
  {Bull}}, \bibinfo {author} {\bibfnamefont {M.~S.}\ \bibnamefont {Scholz}},
  \bibinfo {author} {\bibfnamefont {E.}~\bibnamefont {Carrascosa}}, \bibinfo
  {author} {\bibfnamefont {M.~K.}\ \bibnamefont {Kristiansson}}, \bibinfo
  {author} {\bibfnamefont {G.}~\bibnamefont {Eklund}}, \bibinfo {author}
  {\bibfnamefont {N.}~\bibnamefont {Punnakayathil}}, \bibinfo {author}
  {\bibfnamefont {N.}~\bibnamefont {de~Ruette}}, \bibinfo {author}
  {\bibfnamefont {H.}~\bibnamefont {Zettergren}}, \bibinfo {author}
  {\bibfnamefont {H.~T.}\ \bibnamefont {Schmidt}}, \bibinfo {author}
  {\bibfnamefont {H.}~\bibnamefont {Cederquist}}, \ and\ \bibinfo {author}
  {\bibfnamefont {M.~H.}\ \bibnamefont {Stockett}},\ }\bibfield  {title}
  {\enquote {\bibinfo {title} {Ultraslow radiative cooling of {C}$_n^-$ ($n =$
  3--5)},}\ }\href@noop {} {\bibfield  {journal} {\bibinfo  {journal} {J. Chem.
  Phys.}\ }\textbf {\bibinfo {volume} {151}},\ \bibinfo {pages} {114304}
  (\bibinfo {year} {2019})}\BibitemShut {NoStop}%
\bibitem [{\citenamefont {Bull}\ \emph {et~al.}(2025)\citenamefont {Bull},
  \citenamefont {Subramani}, \citenamefont {Liu}, \citenamefont {Marlton},
  \citenamefont {Ashworth}, \citenamefont {Cederquist}, \citenamefont
  {Zettergren},\ and\ \citenamefont {Stockett}}]{Bull2025}%
  \BibitemOpen
  \bibfield  {author} {\bibinfo {author} {\bibfnamefont {J.~N.}\ \bibnamefont
  {Bull}}, \bibinfo {author} {\bibfnamefont {A.}~\bibnamefont {Subramani}},
  \bibinfo {author} {\bibfnamefont {C.}~\bibnamefont {Liu}}, \bibinfo {author}
  {\bibfnamefont {S.~J.~P.}\ \bibnamefont {Marlton}}, \bibinfo {author}
  {\bibfnamefont {E.~K.}\ \bibnamefont {Ashworth}}, \bibinfo {author}
  {\bibfnamefont {H.}~\bibnamefont {Cederquist}}, \bibinfo {author}
  {\bibfnamefont {H.}~\bibnamefont {Zettergren}}, \ and\ \bibinfo {author}
  {\bibfnamefont {M.~H.}\ \bibnamefont {Stockett}},\ }\bibfield  {title}
  {\enquote {\bibinfo {title} {Radiative cooling in the closed-shell pah cation
  indenyl: A balance betweenrecurrent fluorescence and infrared emission},}\
  }\href@noop {} {\bibfield  {journal} {\bibinfo  {journal} {Under Review}\ }
  (\bibinfo {year} {2025})}\BibitemShut {NoStop}%
\bibitem [{\citenamefont {Iftimie}, \citenamefont {Minary},\ and\ \citenamefont
  {Tuckerman}(2005)}]{Iftimie2005}%
  \BibitemOpen
  \bibfield  {author} {\bibinfo {author} {\bibfnamefont {R.}~\bibnamefont
  {Iftimie}}, \bibinfo {author} {\bibfnamefont {P.}~\bibnamefont {Minary}}, \
  and\ \bibinfo {author} {\bibfnamefont {M.~E.}\ \bibnamefont {Tuckerman}},\
  }\bibfield  {title} {\enquote {\bibinfo {title} {<i>ab initio</i> molecular
  dynamics: Concepts, recent developments, and future trends},}\ }\href
  {\doibase 10.1073/pnas.0500193102} {\bibfield  {journal} {\bibinfo  {journal}
  {Proc. Natl. Acad. Sci.}\ }\textbf {\bibinfo {volume} {102}},\ \bibinfo
  {pages} {6654--6659} (\bibinfo {year} {2005})}\BibitemShut {NoStop}%
\bibitem [{\citenamefont {Dunning}(1989)}]{Dunning1989}%
  \BibitemOpen
  \bibfield  {author} {\bibinfo {author} {\bibfnamefont {T.~H.}\ \bibnamefont
  {Dunning}},\ }\bibfield  {title} {\enquote {\bibinfo {title} {Gaussian basis
  sets for use in correlated molecular calculations. i. the atoms boron through
  neon and hydrogen},}\ }\href {\doibase 10.1063/1.456153} {\bibfield
  {journal} {\bibinfo  {journal} {J. Chem. Phys.}\ }\textbf {\bibinfo {volume}
  {90}},\ \bibinfo {pages} {1007--1023} (\bibinfo {year} {1989})}\BibitemShut
  {NoStop}%
\bibitem [{\citenamefont {Chai}\ and\ \citenamefont
  {Head-Gordon}(2008)}]{Chai2008}%
  \BibitemOpen
  \bibfield  {author} {\bibinfo {author} {\bibfnamefont {J.-D.}\ \bibnamefont
  {Chai}}\ and\ \bibinfo {author} {\bibfnamefont {M.}~\bibnamefont
  {Head-Gordon}},\ }\bibfield  {title} {\enquote {\bibinfo {title} {Long-range
  corrected hybrid density functionals with damped atom–atom dispersion
  corrections},}\ }\href {\doibase 10.1039/b810189b} {\bibfield  {journal}
  {\bibinfo  {journal} {Phys. Chem. Chem. Phys.}\ }\textbf {\bibinfo {volume}
  {10}},\ \bibinfo {pages} {6615} (\bibinfo {year} {2008})}\BibitemShut
  {NoStop}%
\bibitem [{\citenamefont {Bussi}, \citenamefont {Donadio},\ and\ \citenamefont
  {Parrinello}(2007)}]{Bussi2007}%
  \BibitemOpen
  \bibfield  {author} {\bibinfo {author} {\bibfnamefont {G.}~\bibnamefont
  {Bussi}}, \bibinfo {author} {\bibfnamefont {D.}~\bibnamefont {Donadio}}, \
  and\ \bibinfo {author} {\bibfnamefont {M.}~\bibnamefont {Parrinello}},\
  }\bibfield  {title} {\enquote {\bibinfo {title} {Canonical sampling through
  velocity rescaling},}\ }\href {\doibase 10.1063/1.2408420} {\bibfield
  {journal} {\bibinfo  {journal} {J. Chem. Phys.}\ }\textbf {\bibinfo {volume}
  {126}} (\bibinfo {year} {2007}),\ 10.1063/1.2408420}\BibitemShut {NoStop}%
\bibitem [{\citenamefont {Chalyavi}\ \emph {et~al.}(2013)\citenamefont
  {Chalyavi}, \citenamefont {Dryza}, \citenamefont {Sanelli},\ and\
  \citenamefont {Bieske}}]{Chalyavi2013}%
  \BibitemOpen
  \bibfield  {author} {\bibinfo {author} {\bibfnamefont {N.}~\bibnamefont
  {Chalyavi}}, \bibinfo {author} {\bibfnamefont {V.}~\bibnamefont {Dryza}},
  \bibinfo {author} {\bibfnamefont {J.~A.}\ \bibnamefont {Sanelli}}, \ and\
  \bibinfo {author} {\bibfnamefont {E.~J.}\ \bibnamefont {Bieske}},\ }\bibfield
   {title} {\enquote {\bibinfo {title} {Gas-phase electronic spectroscopy of
  the indene cation (c9h8+)},}\ }\href {\doibase 10.1063/1.4808380} {\bibfield
  {journal} {\bibinfo  {journal} {J. Chem. Phys.}\ }\textbf {\bibinfo {volume}
  {138}} (\bibinfo {year} {2013}),\ 10.1063/1.4808380}\BibitemShut {NoStop}%
\bibitem [{\citenamefont {Chu}\ \emph {et~al.}(2023)\citenamefont {Chu},
  \citenamefont {Yu}, \citenamefont {Xiao}, \citenamefont {Zhang},
  \citenamefont {Chen},\ and\ \citenamefont {Zhao}}]{Chu2023}%
  \BibitemOpen
  \bibfield  {author} {\bibinfo {author} {\bibfnamefont {W.}~\bibnamefont
  {Chu}}, \bibinfo {author} {\bibfnamefont {C.}~\bibnamefont {Yu}}, \bibinfo
  {author} {\bibfnamefont {Z.}~\bibnamefont {Xiao}}, \bibinfo {author}
  {\bibfnamefont {Q.}~\bibnamefont {Zhang}}, \bibinfo {author} {\bibfnamefont
  {Y.}~\bibnamefont {Chen}}, \ and\ \bibinfo {author} {\bibfnamefont
  {D.}~\bibnamefont {Zhao}},\ }\bibfield  {title} {\enquote {\bibinfo {title}
  {Gas-phase optical absorption spectra of the indene cation (c <sub>9</sub> h
  <sub>8</sub> <sup>+</sup> )},}\ }\href {\doibase
  10.1080/00268976.2022.2150703} {\bibfield  {journal} {\bibinfo  {journal}
  {Mol. Phys.}\ }\textbf {\bibinfo {volume} {121}} (\bibinfo {year} {2023}),\
  10.1080/00268976.2022.2150703}\BibitemShut {NoStop}%
\bibitem [{\citenamefont {Nagy}\ \emph {et~al.}(2013)\citenamefont {Nagy},
  \citenamefont {Garkusha}, \citenamefont {Fulara},\ and\ \citenamefont
  {Maier}}]{Nagy2013}%
  \BibitemOpen
  \bibfield  {author} {\bibinfo {author} {\bibfnamefont {A.}~\bibnamefont
  {Nagy}}, \bibinfo {author} {\bibfnamefont {I.}~\bibnamefont {Garkusha}},
  \bibinfo {author} {\bibfnamefont {J.}~\bibnamefont {Fulara}}, \ and\ \bibinfo
  {author} {\bibfnamefont {J.~P.}\ \bibnamefont {Maier}},\ }\bibfield  {title}
  {\enquote {\bibinfo {title} {Electronic spectroscopy of transient species in
  solid neon: the indene-motif polycyclic hydrocarbon cation family c9hy+ (y =
  7–9) and their neutrals},}\ }\href {\doibase 10.1039/c3cp52172a} {\bibfield
   {journal} {\bibinfo  {journal} {Phys. Chem. Chem. Phys.}\ }\textbf {\bibinfo
  {volume} {15}},\ \bibinfo {pages} {19091} (\bibinfo {year}
  {2013})}\BibitemShut {NoStop}%
\bibitem [{\citenamefont {West}\ \emph
  {et~al.}(2014{\natexlab{a}})\citenamefont {West}, \citenamefont {Sit},
  \citenamefont {Bodi}, \citenamefont {Hemberger},\ and\ \citenamefont
  {Mayer}}]{West2014}%
  \BibitemOpen
  \bibfield  {author} {\bibinfo {author} {\bibfnamefont {B.}~\bibnamefont
  {West}}, \bibinfo {author} {\bibfnamefont {A.}~\bibnamefont {Sit}}, \bibinfo
  {author} {\bibfnamefont {A.}~\bibnamefont {Bodi}}, \bibinfo {author}
  {\bibfnamefont {P.}~\bibnamefont {Hemberger}}, \ and\ \bibinfo {author}
  {\bibfnamefont {P.~M.}\ \bibnamefont {Mayer}},\ }\bibfield  {title} {\enquote
  {\bibinfo {title} {Dissociative photoionization and threshold photoelectron
  spectra of polycyclic aromatic hydrocarbon fragments: An imaging
  photoelectron photoion coincidence (ipepico) study of four substituted
  benzene radical cations},}\ }\href {\doibase 10.1021/jp5085982} {\bibfield
  {journal} {\bibinfo  {journal} {J. Phys. Chem. A}\ }\textbf {\bibinfo
  {volume} {118}},\ \bibinfo {pages} {11226--11234} (\bibinfo {year}
  {2014}{\natexlab{a}})}\BibitemShut {NoStop}%
\bibitem [{\citenamefont {Klots}(1976)}]{Klots1976}%
  \BibitemOpen
  \bibfield  {author} {\bibinfo {author} {\bibfnamefont {C.~E.}\ \bibnamefont
  {Klots}},\ }\bibfield  {title} {\enquote {\bibinfo {title} {Kinetic energy
  distributions from unimolecular decay: Predictions of the langevin model},}\
  }\href {\doibase 10.1063/1.432111} {\bibfield  {journal} {\bibinfo  {journal}
  {J. Chem. Phys.}\ }\textbf {\bibinfo {volume} {64}},\ \bibinfo {pages}
  {4269--4275} (\bibinfo {year} {1976})}\BibitemShut {NoStop}%
\bibitem [{\citenamefont {Gridelet}\ \emph {et~al.}(2006)\citenamefont
  {Gridelet}, \citenamefont {Lorquet}, \citenamefont {Locht}, \citenamefont
  {Lorquet},\ and\ \citenamefont {Leyh}}]{Gridelet2006}%
  \BibitemOpen
  \bibfield  {author} {\bibinfo {author} {\bibfnamefont {E.}~\bibnamefont
  {Gridelet}}, \bibinfo {author} {\bibfnamefont {A.~J.}\ \bibnamefont
  {Lorquet}}, \bibinfo {author} {\bibfnamefont {R.}~\bibnamefont {Locht}},
  \bibinfo {author} {\bibfnamefont {J.~C.}\ \bibnamefont {Lorquet}}, \ and\
  \bibinfo {author} {\bibfnamefont {B.}~\bibnamefont {Leyh}},\ }\bibfield
  {title} {\enquote {\bibinfo {title} {Hydrogen atom loss from the benzene
  cation. why is the kinetic energy release so large?}}\ }\href {\doibase
  10.1021/jp056119h} {\bibfield  {journal} {\bibinfo  {journal} {J. Phys. Chem.
  A}\ }\textbf {\bibinfo {volume} {110}},\ \bibinfo {pages} {8519--8527}
  (\bibinfo {year} {2006})}\BibitemShut {NoStop}%
\bibitem [{\citenamefont {Hansen}(2018)}]{Hansen2018}%
  \BibitemOpen
  \bibfield  {author} {\bibinfo {author} {\bibfnamefont {K.}~\bibnamefont
  {Hansen}},\ }\bibfield  {title} {\enquote {\bibinfo {title} {Tunneling and
  reflection in unimolecular reaction kinetic energy release distributions},}\
  }\href {\doibase 10.1016/j.cplett.2017.12.075} {\bibfield  {journal}
  {\bibinfo  {journal} {Chem. Phys. Lett.}\ }\textbf {\bibinfo {volume}
  {693}},\ \bibinfo {pages} {66--71} (\bibinfo {year} {2018})}\BibitemShut
  {NoStop}%
\bibitem [{\citenamefont {Laskin}\ and\ \citenamefont
  {Lifshitz}(2001)}]{Laskin2001}%
  \BibitemOpen
  \bibfield  {author} {\bibinfo {author} {\bibfnamefont {J.}~\bibnamefont
  {Laskin}}\ and\ \bibinfo {author} {\bibfnamefont {C.}~\bibnamefont
  {Lifshitz}},\ }\bibfield  {title} {\enquote {\bibinfo {title} {Kinetic energy
  release distributions in mass spectrometry},}\ }\href {\doibase
  10.1002/jms.164} {\bibfield  {journal} {\bibinfo  {journal} {J. Mass
  Spectrom.}\ }\textbf {\bibinfo {volume} {36}},\ \bibinfo {pages} {459--478}
  (\bibinfo {year} {2001})}\BibitemShut {NoStop}%
\bibitem [{\citenamefont {Klots}(1991)}]{Klots1991}%
  \BibitemOpen
  \bibfield  {author} {\bibinfo {author} {\bibfnamefont {C.~E.}\ \bibnamefont
  {Klots}},\ }\bibfield  {title} {\enquote {\bibinfo {title} {Systematics of
  evaporation},}\ }\href {\doibase 10.1007/bf01543949} {\bibfield  {journal}
  {\bibinfo  {journal} {Zeitschrift für Physik D Atoms, Molecules and
  Clusters}\ }\textbf {\bibinfo {volume} {20}},\ \bibinfo {pages} {105--109}
  (\bibinfo {year} {1991})}\BibitemShut {NoStop}%
\bibitem [{\citenamefont {Andersen}, \citenamefont {Bonderup},\ and\
  \citenamefont {Hansen}(2001)}]{Andersen2001}%
  \BibitemOpen
  \bibfield  {author} {\bibinfo {author} {\bibfnamefont {J.}~\bibnamefont
  {Andersen}}, \bibinfo {author} {\bibfnamefont {E.}~\bibnamefont {Bonderup}},
  \ and\ \bibinfo {author} {\bibfnamefont {K.}~\bibnamefont {Hansen}},\
  }\bibfield  {title} {\enquote {\bibinfo {title} {On the concept of
  temperature for a small isolated system},}\ }\href@noop {} {\bibfield
  {journal} {\bibinfo  {journal} {J. Chem. Phys.}\ }\textbf {\bibinfo {volume}
  {114}},\ \bibinfo {pages} {6518--6525} (\bibinfo {year} {2001})}\BibitemShut
  {NoStop}%
\bibitem [{\citenamefont {Andersen}, \citenamefont {Bonderup},\ and\
  \citenamefont {Hansen}(2002)}]{Andersen2002}%
  \BibitemOpen
  \bibfield  {author} {\bibinfo {author} {\bibfnamefont {J.~U.}\ \bibnamefont
  {Andersen}}, \bibinfo {author} {\bibfnamefont {E.}~\bibnamefont {Bonderup}},
  \ and\ \bibinfo {author} {\bibfnamefont {K.}~\bibnamefont {Hansen}},\
  }\bibfield  {title} {\enquote {\bibinfo {title} {Thermionic emission from
  clusters},}\ }\href@noop {} {\bibfield  {journal} {\bibinfo  {journal} {J.
  Phys. B}\ }\textbf {\bibinfo {volume} {35}},\ \bibinfo {pages} {R1--R30}
  (\bibinfo {year} {2002})}\BibitemShut {NoStop}%
\bibitem [{\citenamefont {West}\ \emph
  {et~al.}(2014{\natexlab{b}})\citenamefont {West}, \citenamefont {Joblin},
  \citenamefont {Blanchet}, \citenamefont {Bodi}, \citenamefont {Sztáray},\
  and\ \citenamefont {Mayer}}]{West2014a}%
  \BibitemOpen
  \bibfield  {author} {\bibinfo {author} {\bibfnamefont {B.}~\bibnamefont
  {West}}, \bibinfo {author} {\bibfnamefont {C.}~\bibnamefont {Joblin}},
  \bibinfo {author} {\bibfnamefont {V.}~\bibnamefont {Blanchet}}, \bibinfo
  {author} {\bibfnamefont {A.}~\bibnamefont {Bodi}}, \bibinfo {author}
  {\bibfnamefont {B.}~\bibnamefont {Sztáray}}, \ and\ \bibinfo {author}
  {\bibfnamefont {P.~M.}\ \bibnamefont {Mayer}},\ }\bibfield  {title} {\enquote
  {\bibinfo {title} {Dynamics of hydrogen and methyl radical loss from ionized
  dihydro-polycyclic aromatic hydrocarbons: A tandem mass spectrometry and
  imaging photoelectron–photoion coincidence (ipepico) study of
  dihydronaphthalene and dihydrophenanthrene},}\ }\href {\doibase
  10.1021/jp500430g} {\bibfield  {journal} {\bibinfo  {journal} {J. Phys. Chem.
  A}\ }\textbf {\bibinfo {volume} {118}},\ \bibinfo {pages} {1807--1816}
  (\bibinfo {year} {2014}{\natexlab{b}})},\ \bibinfo {note} {pMID: 24520854},\
  \Eprint {http://arxiv.org/abs/https://doi.org/10.1021/jp500430g}
  {https://doi.org/10.1021/jp500430g} \BibitemShut {NoStop}%
\bibitem [{\citenamefont {West}, \citenamefont {Lowe},\ and\ \citenamefont
  {Mayer}(2018)}]{West2018}%
  \BibitemOpen
  \bibfield  {author} {\bibinfo {author} {\bibfnamefont {B.}~\bibnamefont
  {West}}, \bibinfo {author} {\bibfnamefont {B.}~\bibnamefont {Lowe}}, \ and\
  \bibinfo {author} {\bibfnamefont {P.~M.}\ \bibnamefont {Mayer}},\ }\bibfield
  {title} {\enquote {\bibinfo {title} {Unimolecular dissociation of
  1-methylpyrene cations: Why are 1-methylenepyrene cations formed and not a
  tropylium-containing ion?}}\ }\href {\doibase 10.1021/acs.jpca.8b02667}
  {\bibfield  {journal} {\bibinfo  {journal} {J. Phys. Chem. A}\ }\textbf
  {\bibinfo {volume} {122}},\ \bibinfo {pages} {4730--4735} (\bibinfo {year}
  {2018})},\ \bibinfo {note} {pMID: 29727186},\ \Eprint
  {http://arxiv.org/abs/https://doi.org/10.1021/acs.jpca.8b02667}
  {https://doi.org/10.1021/acs.jpca.8b02667} \BibitemShut {NoStop}%
\bibitem [{\citenamefont {Hashemi}, \citenamefont {Barinovs},\ and\
  \citenamefont {Nyman}(2023)}]{Hashemi2023}%
  \BibitemOpen
  \bibfield  {author} {\bibinfo {author} {\bibfnamefont {S.~R.}\ \bibnamefont
  {Hashemi}}, \bibinfo {author} {\bibfnamefont {G.}~\bibnamefont {Barinovs}}, \
  and\ \bibinfo {author} {\bibfnamefont {G.}~\bibnamefont {Nyman}},\ }\bibfield
   {title} {\enquote {\bibinfo {title} {A reaxff molecular dynamics and rrkm ab
  initio based study on degradation of indene},}\ }\href {\doibase
  10.3389/fspas.2023.1134729} {\bibfield  {journal} {\bibinfo  {journal}
  {Frontiers in Astronomy and Space Sciences}\ }\textbf {\bibinfo {volume}
  {10}} (\bibinfo {year} {2023}),\ 10.3389/fspas.2023.1134729}\BibitemShut
  {NoStop}%
\bibitem [{\citenamefont {Hansen}\ \emph {et~al.}(2001)\citenamefont {Hansen},
  \citenamefont {Andersen}, \citenamefont {Hvelplund}, \citenamefont
  {M{\o}ller}, \citenamefont {Pedersen},\ and\ \citenamefont
  {Petrunin}}]{Hansen2001}%
  \BibitemOpen
  \bibfield  {author} {\bibinfo {author} {\bibfnamefont {K.}~\bibnamefont
  {Hansen}}, \bibinfo {author} {\bibfnamefont {J.~U.}\ \bibnamefont
  {Andersen}}, \bibinfo {author} {\bibfnamefont {P.}~\bibnamefont {Hvelplund}},
  \bibinfo {author} {\bibfnamefont {S.~P.}\ \bibnamefont {M{\o}ller}}, \bibinfo
  {author} {\bibfnamefont {U.~V.}\ \bibnamefont {Pedersen}}, \ and\ \bibinfo
  {author} {\bibfnamefont {V.~V.}\ \bibnamefont {Petrunin}},\ }\bibfield
  {title} {\enquote {\bibinfo {title} {Observation of a 1/t decay law for hot
  clusters and molecules in a storage ring},}\ }\href@noop {} {\bibfield
  {journal} {\bibinfo  {journal} {Phys. Rev. Lett.}\ }\textbf {\bibinfo
  {volume} {87}},\ \bibinfo {pages} {123401} (\bibinfo {year}
  {2001})}\BibitemShut {NoStop}%
\bibitem [{\citenamefont {Barat}\ \emph {et~al.}(2000)\citenamefont {Barat},
  \citenamefont {Brenot}, \citenamefont {Fayeton},\ and\ \citenamefont
  {Picard}}]{Barat2000}%
  \BibitemOpen
  \bibfield  {author} {\bibinfo {author} {\bibfnamefont {M.}~\bibnamefont
  {Barat}}, \bibinfo {author} {\bibfnamefont {J.~C.}\ \bibnamefont {Brenot}},
  \bibinfo {author} {\bibfnamefont {J.~A.}\ \bibnamefont {Fayeton}}, \ and\
  \bibinfo {author} {\bibfnamefont {Y.~J.}\ \bibnamefont {Picard}},\ }\bibfield
   {title} {\enquote {\bibinfo {title} {Absolute detection efficiency of a
  microchannel plate detector for neutral atoms},}\ }\href {\doibase
  10.1063/1.1150615} {\bibfield  {journal} {\bibinfo  {journal} {Rev. Sci.
  Instrum.}\ }\textbf {\bibinfo {volume} {71}},\ \bibinfo {pages} {2050--2052}
  (\bibinfo {year} {2000})},\ \Eprint
  {http://arxiv.org/abs/https://doi.org/10.1063/1.1150615}
  {https://doi.org/10.1063/1.1150615} \BibitemShut {NoStop}%
\bibitem [{\citenamefont {Hansen}(2020)}]{Hansen2020}%
  \BibitemOpen
  \bibfield  {author} {\bibinfo {author} {\bibfnamefont {K.}~\bibnamefont
  {Hansen}},\ }\bibfield  {title} {\enquote {\bibinfo {title} {C$_{60}^-$
  thermal electron-emission rate},}\ }\href {\doibase
  10.1103/physreva.102.052823} {\bibfield  {journal} {\bibinfo  {journal}
  {Phys. Rev. A}\ }\textbf {\bibinfo {volume} {102}},\ \bibinfo {pages}
  {052823} (\bibinfo {year} {2020})}\BibitemShut {NoStop}%
\bibitem [{\citenamefont {Sund\'en}\ \emph {et~al.}(2009)\citenamefont
  {Sund\'en}, \citenamefont {Goto}, \citenamefont {Matsumoto}, \citenamefont
  {Shiromaru}, \citenamefont {Tanuma}, \citenamefont {Azuma}, \citenamefont
  {Andersen}, \citenamefont {Canton},\ and\ \citenamefont
  {Hansen}}]{Sunden2009}%
  \BibitemOpen
  \bibfield  {author} {\bibinfo {author} {\bibfnamefont {A.~E.~K.}\
  \bibnamefont {Sund\'en}}, \bibinfo {author} {\bibfnamefont {M.}~\bibnamefont
  {Goto}}, \bibinfo {author} {\bibfnamefont {J.}~\bibnamefont {Matsumoto}},
  \bibinfo {author} {\bibfnamefont {H.}~\bibnamefont {Shiromaru}}, \bibinfo
  {author} {\bibfnamefont {H.}~\bibnamefont {Tanuma}}, \bibinfo {author}
  {\bibfnamefont {T.}~\bibnamefont {Azuma}}, \bibinfo {author} {\bibfnamefont
  {J.~U.}\ \bibnamefont {Andersen}}, \bibinfo {author} {\bibfnamefont {S.~E.}\
  \bibnamefont {Canton}}, \ and\ \bibinfo {author} {\bibfnamefont
  {K.}~\bibnamefont {Hansen}},\ }\bibfield  {title} {\enquote {\bibinfo {title}
  {Absolute cooling rates of freely decaying fullerenes},}\ }\href {\doibase
  10.1103/PhysRevLett.103.143001} {\bibfield  {journal} {\bibinfo  {journal}
  {Phys. Rev. Lett.}\ }\textbf {\bibinfo {volume} {103}},\ \bibinfo {pages}
  {143001} (\bibinfo {year} {2009})}\BibitemShut {NoStop}%
\bibitem [{\citenamefont {Rasmussen}\ \emph {et~al.}(2022)\citenamefont
  {Rasmussen}, \citenamefont {Teiwes}, \citenamefont {Farkhutdinova},
  \citenamefont {Bochenkova},\ and\ \citenamefont {Andersen}}]{Rasmussen2022}%
  \BibitemOpen
  \bibfield  {author} {\bibinfo {author} {\bibfnamefont {A.~P.}\ \bibnamefont
  {Rasmussen}}, \bibinfo {author} {\bibfnamefont {R.}~\bibnamefont {Teiwes}},
  \bibinfo {author} {\bibfnamefont {D.~A.}\ \bibnamefont {Farkhutdinova}},
  \bibinfo {author} {\bibfnamefont {A.~V.}\ \bibnamefont {Bochenkova}}, \ and\
  \bibinfo {author} {\bibfnamefont {L.~H.}\ \bibnamefont {Andersen}},\
  }\bibfield  {title} {\enquote {\bibinfo {title} {On the temperature of large
  biomolecules in ion-storage rings},}\ }\href {\doibase
  10.1140/epjd/s10053-022-00400-y} {\bibfield  {journal} {\bibinfo  {journal}
  {Eur. Phys. J. D}\ }\textbf {\bibinfo {volume} {76}} (\bibinfo {year}
  {2022}),\ 10.1140/epjd/s10053-022-00400-y}\BibitemShut {NoStop}%
\bibitem [{\citenamefont {Bohme}(1992)}]{Bohme1992}%
  \BibitemOpen
  \bibfield  {author} {\bibinfo {author} {\bibfnamefont {D.~K.}\ \bibnamefont
  {Bohme}},\ }\bibfield  {title} {\enquote {\bibinfo {title} {{PAH}
  [{P}olycyclic {A}romatic {H}ydrocarbons] and fullerene ions and ion/molecule
  reactions in interstellar and circumstellar chemistry},}\ }\href {\doibase
  10.1021/cr00015a002} {\bibfield  {journal} {\bibinfo  {journal} {Chem. Rev.}\
  }\textbf {\bibinfo {volume} {92}},\ \bibinfo {pages} {1487--1508} (\bibinfo
  {year} {1992})},\ \Eprint
  {http://arxiv.org/abs/https://doi.org/10.1021/cr00015a002}
  {https://doi.org/10.1021/cr00015a002} \BibitemShut {NoStop}%
\bibitem [{\citenamefont {Canosa}\ \emph {et~al.}(1995)\citenamefont {Canosa},
  \citenamefont {Laubé}, \citenamefont {Rebrion}, \citenamefont {Pasquerault},
  \citenamefont {Gomet},\ and\ \citenamefont {Rowe}}]{Canosa1995}%
  \BibitemOpen
  \bibfield  {author} {\bibinfo {author} {\bibfnamefont {A.}~\bibnamefont
  {Canosa}}, \bibinfo {author} {\bibfnamefont {S.}~\bibnamefont {Laubé}},
  \bibinfo {author} {\bibfnamefont {C.}~\bibnamefont {Rebrion}}, \bibinfo
  {author} {\bibfnamefont {D.}~\bibnamefont {Pasquerault}}, \bibinfo {author}
  {\bibfnamefont {J.~C.}\ \bibnamefont {Gomet}}, \ and\ \bibinfo {author}
  {\bibfnamefont {B.~R.}\ \bibnamefont {Rowe}},\ }\bibfield  {title} {\enquote
  {\bibinfo {title} {Reaction of anthracene with atomic ions of interstellar
  interest. a falp measurement at room temperature},}\ }\href {\doibase
  10.1016/0009-2614(95)01040-g} {\bibfield  {journal} {\bibinfo  {journal}
  {Chem. Phys. Lett.}\ }\textbf {\bibinfo {volume} {245}},\ \bibinfo {pages}
  {407--414} (\bibinfo {year} {1995})}\BibitemShut {NoStop}%
\bibitem [{\citenamefont {Le~Page}, \citenamefont {Snow},\ and\ \citenamefont
  {Bierbaum}(2001)}]{LePage2001}%
  \BibitemOpen
  \bibfield  {author} {\bibinfo {author} {\bibfnamefont {V.}~\bibnamefont
  {Le~Page}}, \bibinfo {author} {\bibfnamefont {T.~P.}\ \bibnamefont {Snow}}, \
  and\ \bibinfo {author} {\bibfnamefont {V.~M.}\ \bibnamefont {Bierbaum}},\
  }\bibfield  {title} {\enquote {\bibinfo {title} {Hydrogenation and charge
  states of pahspahs in diffuse clouds. i. development of a model},}\ }\href
  {\doibase 10.1086/318952} {\bibfield  {journal} {\bibinfo  {journal} {The
  Astrophysical Journal Supplement Series}\ }\textbf {\bibinfo {volume}
  {132}},\ \bibinfo {pages} {233--251} (\bibinfo {year} {2001})}\BibitemShut
  {NoStop}%
\bibitem [{\citenamefont {Roithová}\ \emph {et~al.}(2006)\citenamefont
  {Roithová}, \citenamefont {Žabka}, \citenamefont {Ascenzi}, \citenamefont
  {Franceschi}, \citenamefont {Ricketts},\ and\ \citenamefont
  {Schröder}}]{Roithova2006}%
  \BibitemOpen
  \bibfield  {author} {\bibinfo {author} {\bibfnamefont {J.}~\bibnamefont
  {Roithová}}, \bibinfo {author} {\bibfnamefont {J.}~\bibnamefont {Žabka}},
  \bibinfo {author} {\bibfnamefont {D.}~\bibnamefont {Ascenzi}}, \bibinfo
  {author} {\bibfnamefont {P.}~\bibnamefont {Franceschi}}, \bibinfo {author}
  {\bibfnamefont {C.~L.}\ \bibnamefont {Ricketts}}, \ and\ \bibinfo {author}
  {\bibfnamefont {D.}~\bibnamefont {Schröder}},\ }\bibfield  {title} {\enquote
  {\bibinfo {title} {Energetics of fragmentations of indene dication from
  photoionization experiments},}\ }\href {\doibase
  10.1016/j.cplett.2006.03.083} {\bibfield  {journal} {\bibinfo  {journal}
  {Chem. Phys. Lett.}\ }\textbf {\bibinfo {volume} {423}},\ \bibinfo {pages}
  {254--259} (\bibinfo {year} {2006})}\BibitemShut {NoStop}%
\bibitem [{\citenamefont {Snow}\ \emph {et~al.}(1998)\citenamefont {Snow},
  \citenamefont {Page}, \citenamefont {Keheyan},\ and\ \citenamefont
  {Bierbaum}}]{Snow1998}%
  \BibitemOpen
  \bibfield  {author} {\bibinfo {author} {\bibfnamefont {T.~P.}\ \bibnamefont
  {Snow}}, \bibinfo {author} {\bibfnamefont {V.~L.}\ \bibnamefont {Page}},
  \bibinfo {author} {\bibfnamefont {Y.}~\bibnamefont {Keheyan}}, \ and\
  \bibinfo {author} {\bibfnamefont {V.~M.}\ \bibnamefont {Bierbaum}},\
  }\bibfield  {title} {\enquote {\bibinfo {title} {The interstellar chemistry
  of {PAH} cations},}\ }\href {\doibase 10.1038/34602} {\bibfield  {journal}
  {\bibinfo  {journal} {Nature}\ }\textbf {\bibinfo {volume} {391}},\ \bibinfo
  {pages} {259--260} (\bibinfo {year} {1998})}\BibitemShut {NoStop}%
\bibitem [{\citenamefont {Betts}\ \emph {et~al.}(2006)\citenamefont {Betts},
  \citenamefont {Stepanovic}, \citenamefont {Snow},\ and\ \citenamefont
  {Bierbaum}}]{Betts2006}%
  \BibitemOpen
  \bibfield  {author} {\bibinfo {author} {\bibfnamefont {N.~B.}\ \bibnamefont
  {Betts}}, \bibinfo {author} {\bibfnamefont {M.}~\bibnamefont {Stepanovic}},
  \bibinfo {author} {\bibfnamefont {T.~P.}\ \bibnamefont {Snow}}, \ and\
  \bibinfo {author} {\bibfnamefont {V.~M.}\ \bibnamefont {Bierbaum}},\
  }\bibfield  {title} {\enquote {\bibinfo {title} {Gas-phase study of coronene
  cation reactivity of interstellar relevance},}\ }\href {\doibase
  10.1086/509875} {\bibfield  {journal} {\bibinfo  {journal} {Astrophys. J.}\
  }\textbf {\bibinfo {volume} {651}},\ \bibinfo {pages} {L129--L131} (\bibinfo
  {year} {2006})}\BibitemShut {NoStop}%
\bibitem [{\citenamefont {Wakelam}\ and\ \citenamefont
  {Herbst}(2008)}]{Wakelam2008}%
  \BibitemOpen
  \bibfield  {author} {\bibinfo {author} {\bibfnamefont {V.}~\bibnamefont
  {Wakelam}}\ and\ \bibinfo {author} {\bibfnamefont {E.}~\bibnamefont
  {Herbst}},\ }\bibfield  {title} {\enquote {\bibinfo {title} {Polycyclic
  aromatic hydrocarbons in dense cloud chemistry},}\ }\href {\doibase
  10.1086/587734} {\bibfield  {journal} {\bibinfo  {journal} {Astrophys. J.}\
  }\textbf {\bibinfo {volume} {680}},\ \bibinfo {pages} {371--383} (\bibinfo
  {year} {2008})}\BibitemShut {NoStop}%
\bibitem [{\citenamefont {Lepp}\ and\ \citenamefont
  {Dalgarno}(1988)}]{Lepp1988}%
  \BibitemOpen
  \bibfield  {author} {\bibinfo {author} {\bibfnamefont {S.}~\bibnamefont
  {Lepp}}\ and\ \bibinfo {author} {\bibfnamefont {A.}~\bibnamefont
  {Dalgarno}},\ }\bibfield  {title} {\enquote {\bibinfo {title} {{P}olycyclic
  {A}romatic {H}ydrocarbons in interstellar chemistry},}\ }\href {\doibase
  10.1086/165915} {\bibfield  {journal} {\bibinfo  {journal} {Astrophys. J.}\
  }\textbf {\bibinfo {volume} {324}},\ \bibinfo {pages} {553} (\bibinfo {year}
  {1988})}\BibitemShut {NoStop}%
\end{thebibliography}
\end{document}